\documentclass[journal,12pt,onecolumn]{IEEEtran}
\usepackage{bbm, enumitem, indentfirst}
\usepackage{amsmath, amssymb, amsthm}  
\usepackage{rotating, booktabs}  
\usepackage{caption, graphicx, array, subfigure, url, color}
\usepackage[ruled]{algorithm2e}
\usepackage{algpseudocode}

\SetKwRepeat{Do}{do}{while}

\newcommand {\mymarginpar}[1]{\marginpar{#1}}
\renewcommand {\marginpar}[1]{}

\def\_{\rule{.3em}{.15ex}}      

\newcommand{\ls}[1]
   {\dimen0=\fontdimen6\the\font
    \lineskip=#1\dimen0
    \advance\lineskip.5\fontdimen5\the\font
    \advance\lineskip-\dimen0
    \lineskiplimit=.9\lineskip
    \baselineskip=\lineskip
    \advance\baselineskip\dimen0
    \normallineskip\lineskip
    \normallineskiplimit\lineskiplimit
    \normalbaselineskip\baselineskip
    \ignorespaces
   }

\newtheorem{definition}{Definition}
\newtheorem{property}[definition]{Property}
\newtheorem{proposition}[definition]{Proposition}
\newtheorem{lemma}[definition]{Lemma}
\newtheorem{theorem}[definition]{Theorem}
\newtheorem{corollary}[definition]{Corollary}
\newtheorem{example}[definition]{Example}
\newtheorem{remark}[definition]{Remark}

\newcommand {\benum} {\begin{enumerate}}
\newcommand {\eenum} {\end{enumerate}}
\newcommand {\bdesc} {\begin{description}}
\newcommand {\edesc} {\end{description}}

\newcommand {\bfig}[2] {\begin{figure}
  \centering
  \includegraphics[width=#2]{#1}}
\newcommand {\brotatefig}[2] {\begin{figure}[htbp]
                        \centerline {
                         \epsfig{figure={#1},clip=,angle=-90,width={#2}}}}

\newcommand {\bfigfirst}[2] {\begin{figure}[h]
                        \centerline {
                        \setlength{\epsfxsize}{#2}
                        \epsffile{#1}}}
\newcommand {\efig}[2]{ \caption{#2}
                        \label{fig:#1}
                        \end{figure}
                        \mymarginpar{fig:#1}}
\newcommand {\erotatefig}[2]{ \caption{#2}
                        \label{fig:#1}
                        \end{figure}
                        \mymarginpar{fig:#1}}
\newcommand {\rfig}[1]{Figure \ref{fig:#1}}

\newcommand {\btab}[1]{
                       \begin{table}
                       \centering
                       \begin{tabular}{#1}}
\newcommand {\etab}[3] {
                       \end{tabular}
                       \caption[#3]{#2}
                       \label{tab:#1}
                       \end{table}
                       \mymarginpar{tab:#1}
                       \vspace{.1in}}

\newcommand {\btabular}[1]{\begin{center}
                       \begin{tabular}{#1}}
\newcommand {\etabular}{\end{tabular}
                       \end{center}}

\newcommand {\bdefin}[1]{\begin{definition}
                      \mymarginpar{def:#1}
                      \label{def:#1} }
\newcommand {\edefin}       {\end{definition}}
\newcommand {\rdef}[1]{Definition \ref{def:#1}}

\newcommand {\bpro}[1]{\begin{property}
                      \mymarginpar{pro:#1}
                      \label{pro:#1} }
\newcommand {\epro}   {\end{property}}

\newcommand {\bprop}[1]{\begin{proposition}
                      \mymarginpar{prop:#1}
                      \label{prop:#1} }
\newcommand {\eprop}       {\end{proposition}}

\newcommand {\blem}[1]{\begin{lemma}
                      \mymarginpar{lem:#1}
                      \label{lem:#1} }
\newcommand {\elem}   {\end{lemma}}
\newcommand {\rlem}[1]{Lemma \ref{lem:#1}}

\newcommand {\bthe}[1]{\begin{theorem}
                      \mymarginpar{the:#1}
                      \label{the:#1} }
\newcommand {\ethe}   {\end{theorem}}
\newcommand {\rthe}[1]{Theorem \ref{the:#1}}

\newcommand {\bproof}{\noindent {\bf Proof.} \ }
\newcommand {\eproof} {\hfill \squares \\ \vspace{.3cm}}

\newcommand {\bcor}[1]{\begin{corollary}
                      \mymarginpar{cor:#1}
                      \label{cor:#1} }
\newcommand {\ecor}   {\end{corollary}}
\newcommand {\rcor}[1]{Corollary \ref{cor:#1}}

\newcommand {\bax}[1]{\begin{axiom}
                      \mymarginpar{ax:#1}
                      \label{ax:#1} }
\newcommand {\eax}       {\vspace{-.1in} \end{axiom}}

\newcommand {\bex}[2]{\vspace{.1in}
                      \begin{example}
                      \mymarginpar{ex:#1}
                       {\bf #2}
                      \label{ex:#1} }
\newcommand {\eex}       {\end{example} \vspace{.3cm} }

\newcommand {\brem}[1]{\begin{remark}
                      \mymarginpar{rem:#1}
                      \label{rem:#1} \em }
\newcommand {\erem}   {\end{remark}}

\newcommand {\beq}[1]{\mymarginpar{eq:#1}
                      \begin{equation}
                      \label{eq:#1} }

\newcommand {\beqno}[1]{\mymarginpar{eq:#1}
                      \begin{eqnarray}
                      \nonumber}

\newcommand {\eeq}       {\end{equation}}
\newcommand {\eeqno}       { && \end{eqnarray}}
\newcommand {\req}[1]{(\ref{eq:#1})}

\newcommand {\bear}[1]{\mymarginpar{eq:#1}
                       \begin{eqnarray}
                       \label{eq:#1} }

\newcommand {\bearno}[1]{\mymarginpar{eq:#1}
                       \begin{eqnarray}
                       \nonumber}

\newcommand {\eear}{\end{eqnarray}}
\newcommand {\eearno}{\end{eqnarray}}

\newcommand {\bsel}{\left \{ \begin{array}{cl}}
\newcommand {\esel}{\end{array} \right.}

\newcommand {\bmat}[1]{\left [ \begin{array}{#1}}
\newcommand {\emat}{\end{array} \right ]}

\def\R{I\kern-0.30em R}
\def\N{I\kern-0.30em N}
\def\P{I\kern-0.30em P}

\def\ex{{\bf\sf E}}

\newcommand\squares{\vrule height6pt width7pt depth1pt}

\newcommand\Deltaw{\Delta_{\mbox{a}}}

\newcommand\ksetsplus{$\mbox{K-sets}^+$ }
\newcommand\gsim{\gamma}

\IEEEoverridecommandlockouts

\begin{document}

\title{A Unified Framework for Sampling, Clustering and Embedding Data Points in Semi-Metric Spaces}

\date{July, 2017}

\author{Chia-Tai~Chang and~Cheng-Shang~Chang

\thanks{
C.-T. Chang and C.-S. Chang are with the Institute of Communications Engineering,
National Tsing Hua University, Hsinchu 30013, Taiwan, R.O.C.
Email: s104064540@m104.nthu.edu.tw; cschang@ee.nthu.edu.tw.}}

\maketitle

\begin{abstract}

In this paper, we propose a unified framework for sampling, clustering and embedding data points in semi-metric spaces. For a set of data points $\Omega=\{x_1, x_2, \ldots, x_n\}$ in a semi-metric space, there is a semi-metric that measures the distance between two points. Our idea of sampling the data points in a semi-metric space is to consider a complete graph with $n$ nodes and $n$ self edges and then map each data point in $\Omega$ to a node in the graph with the edge weight between two nodes being the distance between the corresponding two points in $\Omega$. By doing so, several well-known sampling techniques developed for community detections in graphs can be applied for clustering data points in a semi-metric space. One particularly interesting sampling technique is the exponentially twisted sampling in which one can specify the desired average distance from the sampling distribution to detect clusters with various resolutions. 

Each sampling distribution leads to a covariance matrix that measures how two points are correlated. By using a covariance matrix as input, we also propose a softmax clustering algorithm that can be used for not only clustering but also embedding data points in a semi-metric space to a low dimensional Euclidean space. Our experimental results show that after a certain number of iterations of ``training,'' our softmax algorithm can reveal the ``topology'' of the data from a high dimensional Euclidean space by only using the pairwise distances. To provide further theoretical support for our findings, we show that the eigendecomposition of a covariance matrix is equivalent to the principal component analysis (PCA) when the squared Euclidean distance is used as the semi-metric for high dimensional data. 

To deal with the hierarchical structure of clusters, our softmax clustering algorithm can also be used with a hierarchical agglomerative clustering algorithm. For this, we propose an iterative partitional-hierarchical algorithm, called $i$PHD, in this paper. Both the softmax clustering algorithm and the $i$PHD algorithm are modularity maximization algorithms. On the other hand, the $K$-means algorithm and the $K$-sets algorithm in the literature are based on maximization of the normalized modularity. We compare our algorithms with these existing algorithms to show how the choice of the objective function and the choice of the distance measure affect the performance of the clustering results. Our experimental results show that those algorithms based on the maximization of normalized modularity tend to balance the sizes of detected clusters and thus do not perform well when the ground-truth clusters are different in sizes. Also, using  a metric is better than using a semi-metric as the triangular inequality is not satisfied for a semi-metric and that is more prone to clustering errors.

\end{abstract}


\IEEEpeerreviewmaketitle

\section{Introduction}

\IEEEPARstart{T}{he} community detection/clustering is a fundamental technology for data analysis and it has a lot of applications in various fields, including machine learning, social network analysis, and computational biology for protein sequences. It is worth noting that clustering is in general considered as an ill-posed problem, but previous studies always claim that there exist a ground-truth of clusters.

\begin{figure}[htbp]
	\centering
	\begin{tabular}{p{0.22\textwidth}p{0.22\textwidth}p{0.22\textwidth}p{0.22\textwidth}}
		\includegraphics[width=0.22\textwidth]{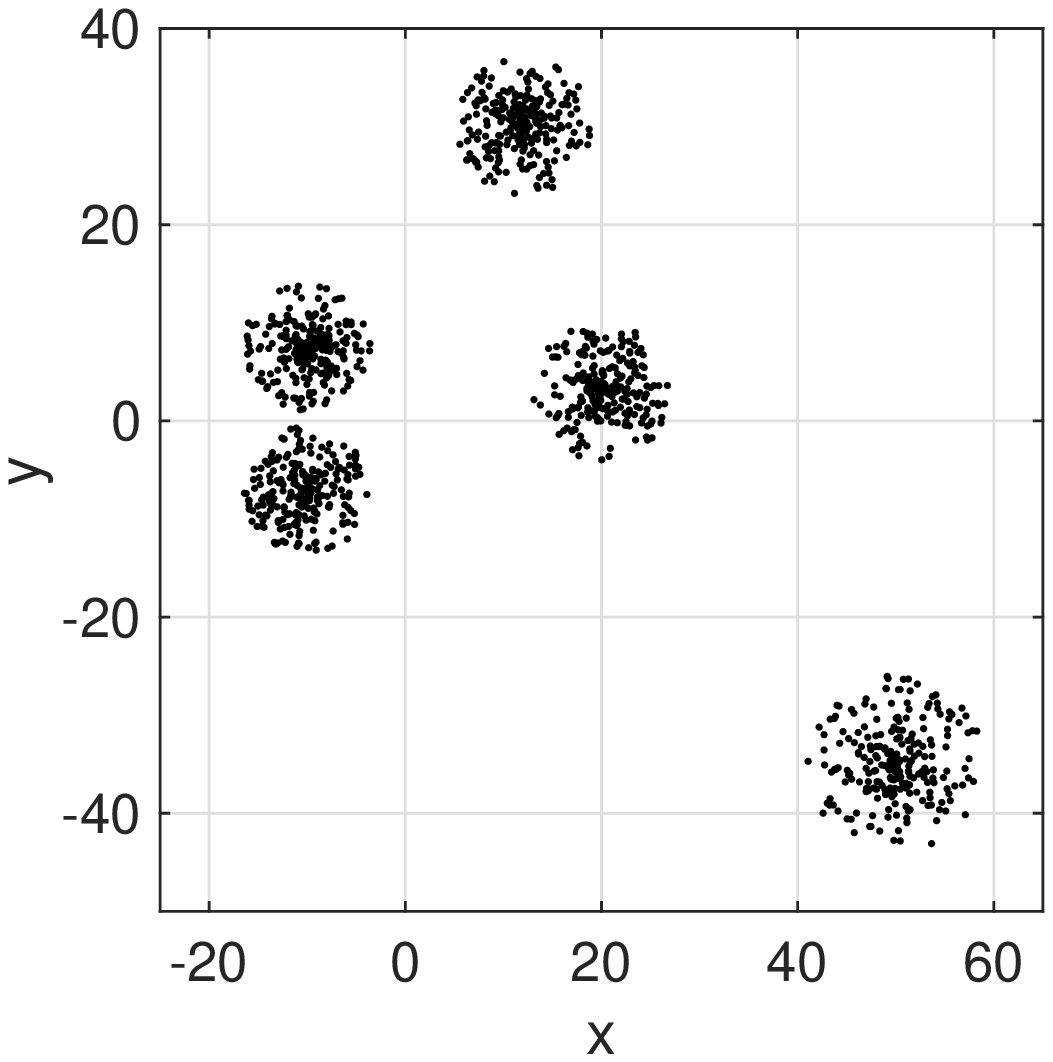} &
		\includegraphics[width=0.22\textwidth]{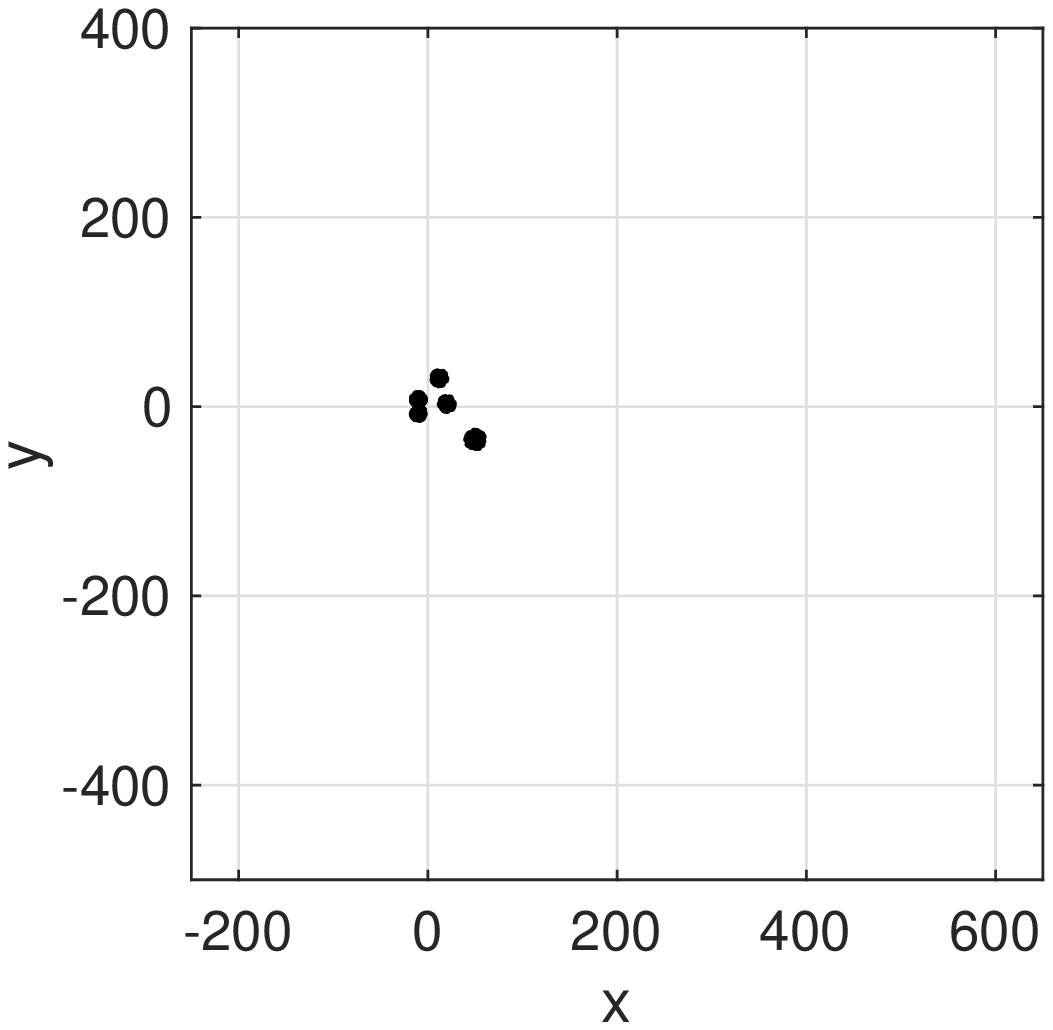} &
        \includegraphics[width=0.22\textwidth]{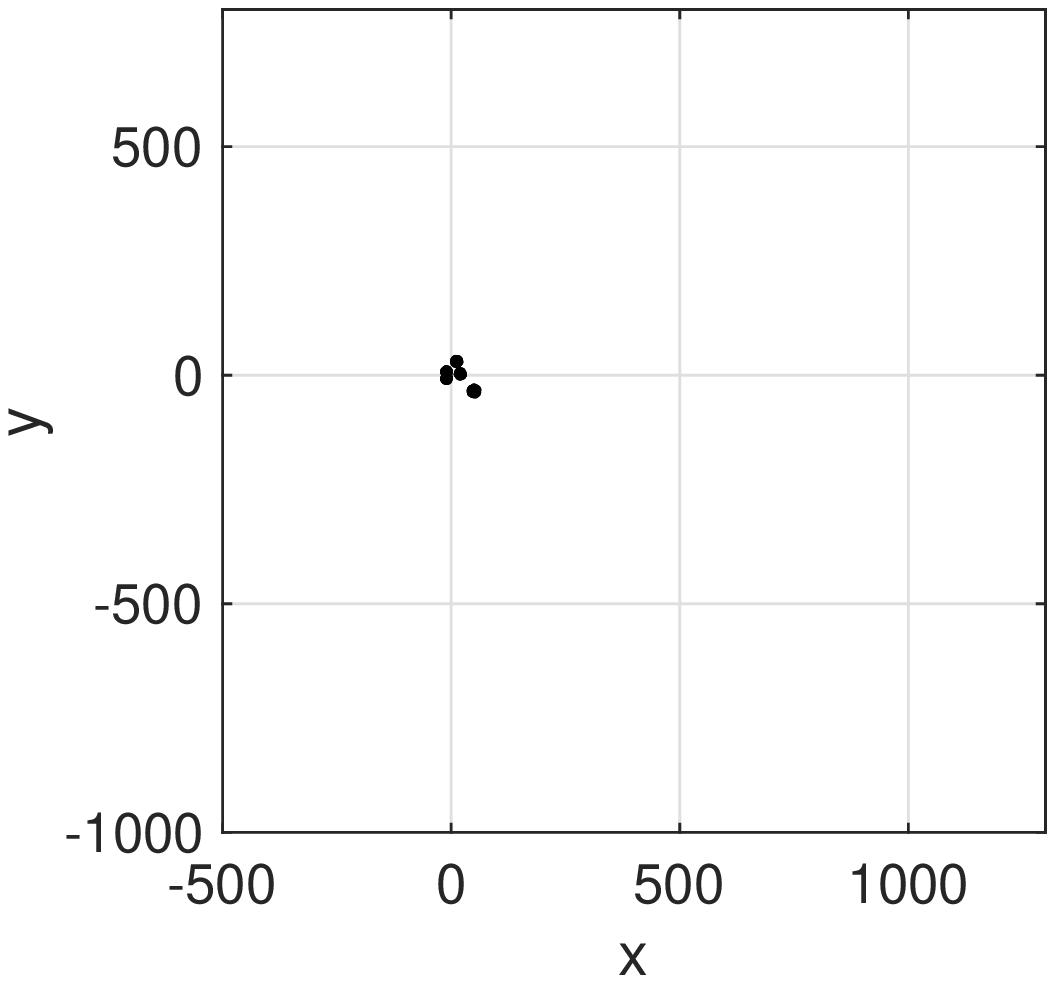} &
		\includegraphics[width=0.22\textwidth]{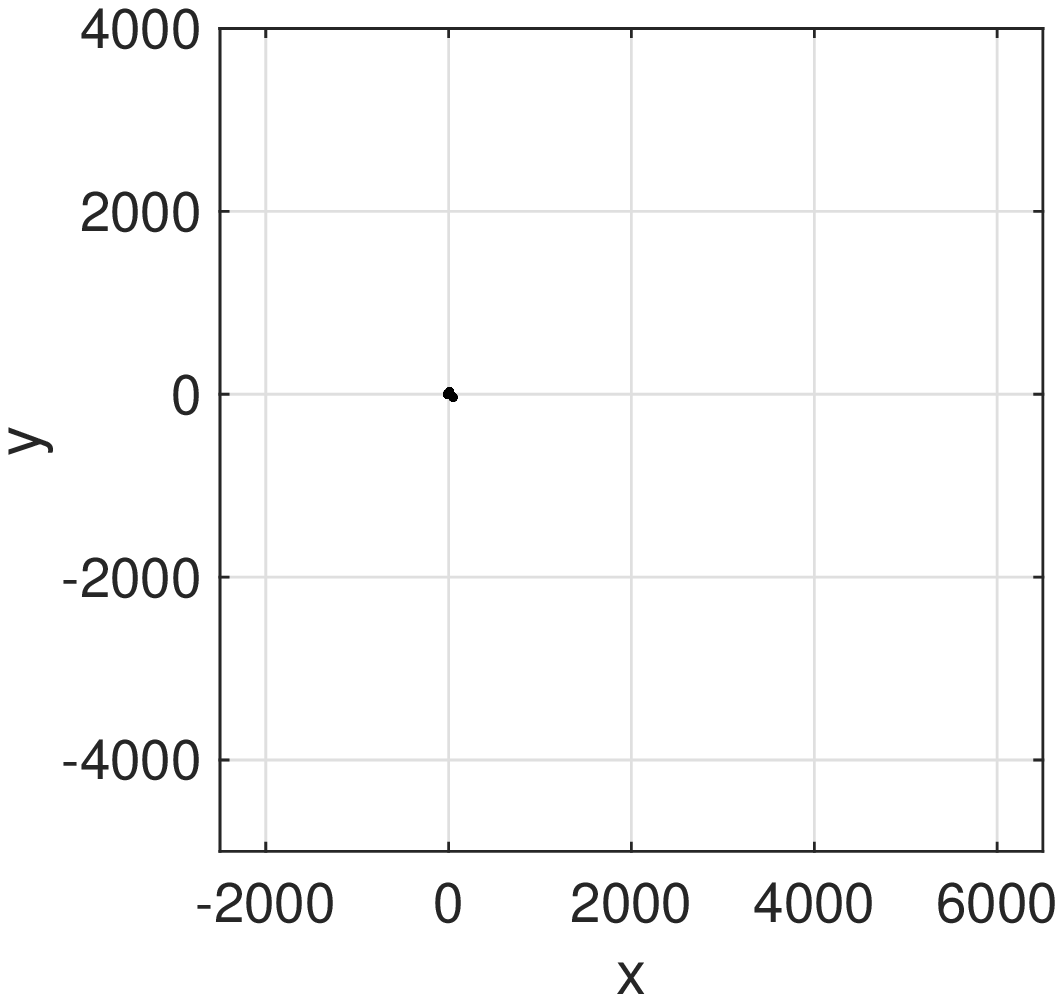} \\
		(a) Resolution 1 & (b) Resolution 2 & (c) Resolution 3 & (d) Resolution 4 \\
	\end{tabular}
	\caption{A view at different resolutions.}
	\label{fig:DiffResolution}
\end{figure}

In \rfig{DiffResolution}, for example, one takes a quick glance at (a) will say that there are five communities in the two-dimensional space and that is more than the others have. In practically, those figures are exactly the same dataset but present at different resolutions. We can say that such a community detection/clustering problem do not have a unique ground-truth, that is, the answer depends on your perspectives. For this, we developed a probabilistic-based framework with the adaptive parameters. The key idea of the framework is to sample a network by the exponentially twisted sampling which helps us to detect the graph with various resolution.

One of our main contributions of this paper is that we provide a unified framework for sampling, clustering, and embedding in semi-metric spaces, where the distance measure does not necessarily satisfy the triangular inequality. Unlike most previous methods, by using a covariance matrix as input, the softmax clustering algorithm that can be used for not only clustering but also embedding data points in a semi-metric space to a low dimensional Euclidean space. Also, we show that the eigendecomposition of a covariance matrix is equivalent to the principal component analysis (PCA) when the squared Euclidean distance is used as the semi-metric for high dimensional data. However, there are two drawbacks of the softmax clustering algorithm: (i) the output of the algorithm may not be a cluster, and (ii) the dataset may exist a hierarchical structure of clusters. For this, we introduce the iterative Partitional-Hierarchical community Detection ($i$PHD) algorithm, and our idea is to add a hierarchical agglomerative clustering algorithm after the softmax clustering algorithm. We then show an illustration for various resolutions with different average distances $\bar{d}$ by selecting various $\lambda$.

Another contribution of this paper is to discuss the performance comparison problem between algorithms. We can divide the problem into two parts: (i) choice of the objective function, and (ii) choice of the distance measure affect the performance of the clustering results. In the first part, both algorithms we propose above are modularity maximization algorithms. On the other hand, the $K$-means algorithm and the $K$-sets algorithm in the literature are based on maximization of the normalized modularity. To evaluate the performance of $i$PHD and $K$-sets$^+$ algorithm, we conduct two experiments: (i) community detection of signed networks generated by the stochastic block model, and (ii) clustering of a real network from the LiveJournal dataset \cite{LiveJournal1,LiveJournal2}. Our experiments show that the normalized modularity tends to balance the sizes of the detected communities. Besides, there is an interesting situation that the wrong clustering result has a higher objective value. For this, via the duality between a semi-cohesion and a semi-metric, and the shortest-path algorithm, we can convert a similarity measure into a distance metric then clustering by the $K$-sets algorithm. The experimental results show that using a metric is better than using a semi-metric as the triangular inequality in not satisfied for a semi-metric and that might cause misclustering of some data points.

The rest of this paper is organized as follows. In Section 2, we introduce a probabilistic framework of graph sampling and propose the softmax clustering algorithm and the iterative Partitional-Hierarchical community Detection ($i$PHD) algorithm. In Section 3, we discuss the performance comparison problem between algorithms and the pros and cons of the choice of the distance measures. The experimental results are presented after each corresponding section. The paper is concluded in Section 4.

\section{Clustering and embedding data points in semi-metric spaces}

\subsection{Semi-metrics and semi-cohesion measures}

In this section, we consider the  problem of embedding data points in a semi-metric space to a low dimensional Euclidean space. For this, we consider a set of $n$ data points, $\Omega=\{x_1, x_2, \ldots, x_n\}$ and a distance measure $d(x,y)$ for any two points $x$ and $y$ in $\Omega$. The distance measure $d(\cdot, \cdot)$  is assumed to be a {\em semi-metric} and it satisfies the following three properties:
\begin{description}
\item[(D1)] (Nonnegativity) $d(x,y) \ge 0$.
\item[(D2)] (Null condition) $d(x,x)=0$.
\item[(D3)] (Symmetry) $d(x,y)=d(y,x)$.
\end{description}

The semi-metric assumption is weaker than the metric assumption in \cite{ksets}, where the distance measure is assumed to satisfy the triangular inequality.

Given a semi-metric $d(\cdot,\cdot)$  for $\Omega$, it is defined in \cite{ksetsplus} the induced semi-cohesion measure as follows:
\bear{csim7777}
\gsim(x,y)&=&\frac{1}{n}\sum_{z_2 \in \Omega} d(z_2,y)+\frac{1}{n} \sum_{z_1 \in \Omega}d(x,z_1)\nonumber\\
&-&\frac{1}{n^2} \sum_{z_2 \in \Omega}\sum_{z_1 \in \Omega}d(z_2,z_1)-d(x,y).
\eear
It was shown in \cite{ksetsplus} that the induced semi-cohesion measure
satisfies
the following three properties:
\begin{description}
\item[(C1)] (Symmetry) $\gsim(x,y)=\gsim(y,x)$ for all $x, y \in \Omega$.
\item[(C2)] (Null condition) For all $x \in \Omega$, $\sum_{y \in \Omega}\gsim (x,y)=0$.
\item[(C3)] (Nonnegativity) For all $x, y$ in $\Omega$,
\beq{cmeas1111}
\gsim(x,x)+\gsim (y,y) \ge 2\gsim (x, y).
\eeq
\end{description}
Moreover, one also has
\beq{cind1111}
d(x,y)=(\gsim(x,x)+\gsim(y,y))/2 -\gsim(x,y).
\eeq
Thus, there is a one-to-one mapping (a duality result) between a semi-metric and a semi-cohesion measure. In this paper, we will simply say data points are in a semi-metric space if there is either a semi-cohesion measure or a semi-metric associated with these data points.

The embedding problem is usually to map the data points in a semi-metric space to a low dimensional Euclidean space so that the change of the distance between any two points can be minimized. In general, such an embedding problem is formulated as an optimization problem that minimizes the sum of the errors of the distances and thus requires a very high computational effort. Instead of preserving the distances, our objective in this paper is merely to preserve the ``topology'' of the data points. Our idea of doing this is to use a self-organized clustering algorithm to cluster data points into $K$ sets and then embed each data point into a $K$-dimensional Euclidean space by using its covariance to the $K$ clusters as the $K$ coordinates.

\subsection{Exponentially twisted sampling}

In this section, we use the notion of sampled graphs in \cite{chang2013relative,chang2016} to define the covariance/correlation measure. For this set of data points in a semi-metric space, we can view it as a complete graph with $n$ self edges and an edge between two points is assigned with a distance measure. If we sample two points $X$and $Y$ {\em uniformly}, then we have the following sampling bivariate distribution:
\beq{exp1111}
p(x,y)=\frac{1}{n^2},
\eeq
for all $x,y \in\Omega$.
Using such a sampling distribution, the average distance between two randomly selected points is
\beq{exp2222}
\ex_p[d(X,Y)]=\frac{1}{n^2}\sum_{x \in \Omega}\sum_{y \in \Omega}d(x,y).
\eeq
Suppose we would like to change another sampling distribution $p_\lambda(x,y)$ so that the average distance between two randomly selected points, denoted by ${\bar d}$, is smaller than $\ex_p[d(X,Y)]$ in \req{exp2222}. By doing so, a pair of two points with a shorter distance is selected more often than another pair of two points with a larger distance. For this, we consider the following
minimization problem:
\bear{exp3333}
&\min \quad &D(p_\lambda \Vert p)\nonumber \\
&s.t. \quad &\sum_{x \in \Omega}\sum_{y \in \Omega}p_\lambda (x,y)=1, \nonumber \\
&&\sum_{x \in \Omega}\sum_{y \in \Omega}d(x,y) \cdot p_\lambda (x,y)={\bar d},
\eear
 where $D(p_\lambda \Vert p)$ is the
 Kullback-Leibler distance between the two probability mass functions $p_\lambda(x,y)$ and $p(x,y)$, i.e.,
\beq{exp4444}
D(p_\lambda \Vert p)=\sum_{x \in \Omega}\sum_{y \in \Omega}p_\lambda (x,y) \log (\frac{p_\lambda (x,y)}{p(x,y)}).
\eeq
The solution of such a minimization problem is known to be the exponentially twisted distribution as follows:
\beq{exp5555}
p_\lambda(x,y)=C*\exp(\lambda \cdot d(x,y))*p(x,y),
\eeq
where
\beq{exp6666}
C=\frac{1}{\sum_{x \in \Omega}\sum_{y \in \Omega}\exp(\lambda \cdot d(x,y))*p(x,y)}
\eeq
is the normalization constant. As the distance measure $d(\cdot,\cdot)$ is symmetric and $p(x,y)=1/n^2$, we know that $p_{\lambda}(x,y)=p_{\lambda}(y,x)$.
Moreover, the parameter $\lambda$ can be solved by the following equation:
\beq{exp7777}
\frac{\partial F}{\partial \lambda}=\sum_{x \in \Omega}\sum_{y \in \Omega}d(x,y) \cdot p_\lambda (x,y)={\bar d} ,
\eeq
where $F=\log(1/C)$ is the energy function.

\begin{figure}[htbp]
	\centering
	\begin{tabular}{p{0.45\textwidth}p{0.45\textwidth}}
		\includegraphics[width=0.45\textwidth]{BlackGT.eps} &
		\includegraphics[width=0.45\textwidth]{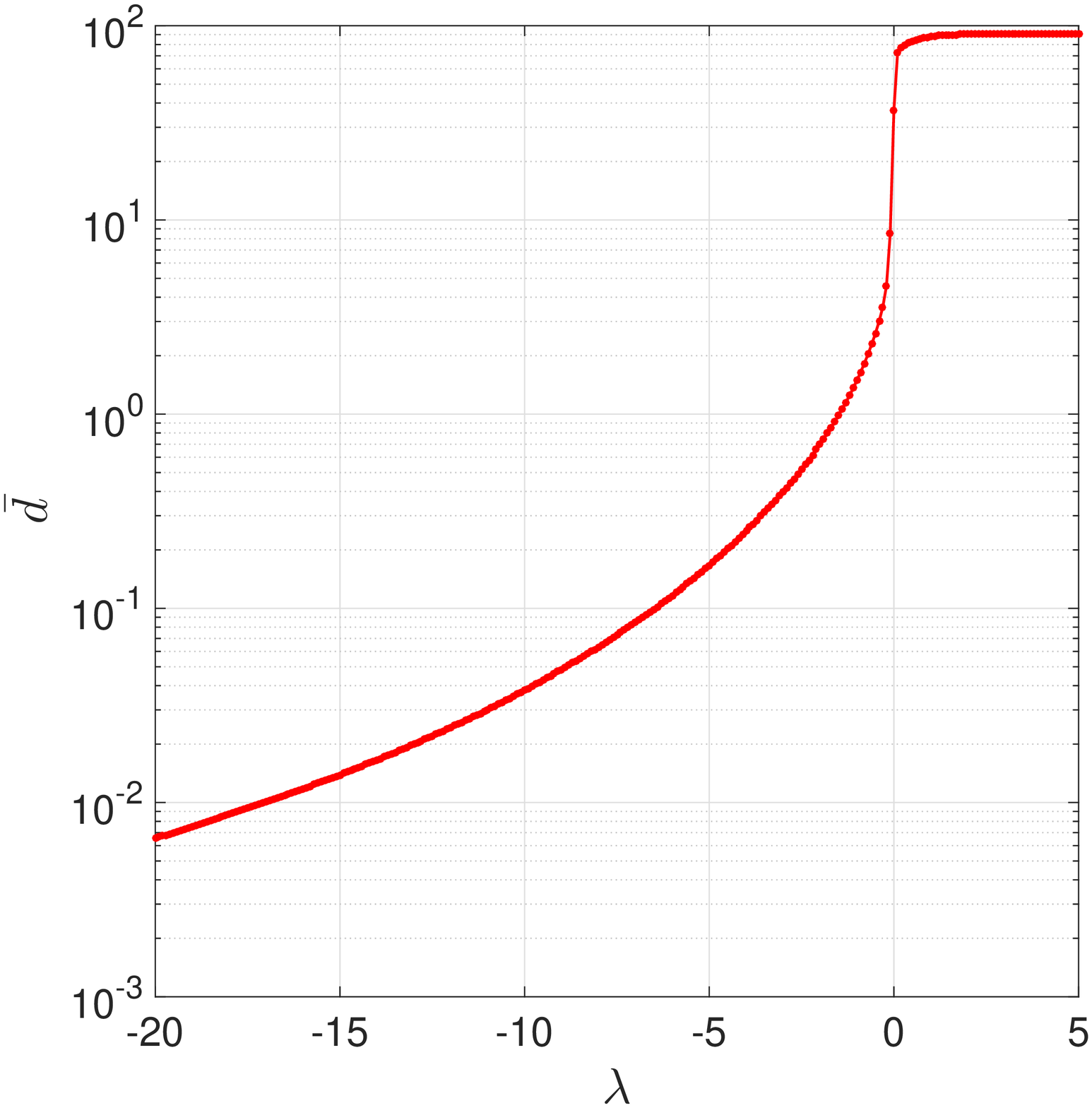} \\
		(a) The 2D plot. & (b) The parameter $\bar d$ as a function of $\lambda$.\\
	\end{tabular}
	\caption{The five-circle dataset.}
	\label{fig:fivecircle}
\end{figure}

If we choose $\lambda<0$, then ${\bar d}\le \ex_p[d(X,Y)]$. To illustrate this, we generate a five-circle dataset on a plane (as shown in \rfig{fivecircle} (a)), where each circle contains 250 points and thus a total number of 1250 points in this dataset. In \rfig{fivecircle} (b), we plot the average distance $\bar d$ as a function of $\lambda$. Clearly, as $\lambda \to \infty$, $\bar d$ approaches to the maximum distance between a pair of two points. On the other hand, as $\lambda \to -\infty$, $\bar d$ approaches to the minimum distance between a pair of two points. The plot in \rfig{fivecircle} (b) allows us to solve \req{exp7777} numerically.

\subsection{Clusters in a sampled graph}

Now with the exponentially twisted distribution, we can define sampled graphs as in \cite{chang2013relative,chang2016}.

\bdefin{sampled}{\bf (Sampled graph \cite{chang2016})} Let $G$ be the complete graph with $n$ nodes and $n$ self edges. Each node in $G$ corresponds to a specific data point in $\Omega$ and the edge weight between two nodes is the distance measure between the corresponding two points in $\Omega$. The graph $G$ sampled by randomly selecting an ordered pair of two nodes $(X,Y)$ according to the symmetric bivariate distribution $p_{\lambda}(x,y)$  in \req{exp5555} is called a {\em sampled graph} and it is denoted by the two-tuple $(G,p_\lambda(\cdot,\cdot))$.
\edefin

Let
\beq{exp7711}
p_{\lambda}(x)=\sum_{y \in \Omega}p_{\lambda}(x,y)
\eeq
be the marginal probability that the point $x$ is selected by using the sampling distribution $p_\lambda(\cdot,\cdot)$.
The higher the probability is, the more important that point is. As such, $p_{\lambda}(x)$ can be used for ranking data points according to
the sampling distribution $p_\lambda(\cdot,\cdot)$, and it is called the {\em centrality} of point $x$ for the sampled graph $(G,p_\lambda(\cdot,\cdot))$. For example, if we use the original sampling distribution (with $\lambda=0$), then
\beq{exp7722}
p_0(x)=\frac{1}{n^2}\sum_{y \in \Omega}d(x,y)
\eeq
and it is known as the {\em closeness centrality} in the literature (see e.g., the book \cite{Newman2010}).
In this paper, we use $C(x)$ to denote the centrality of $x$, i.e., $C(x)=p_\lambda(x)$, and $C(S)$ to denote the centrality of the points in $S$, i.e.,
$C(S)=\sum_{x \in S}C(x)$.
As a generalization of  centrality, relative centrality in \cite{chang2013relative} is a (probability) measure that measures how important a set of nodes in a network is with respect to another set of nodes.

\bdefin{relativem} {\bf (Relative centrality)}
For a sampled graph $(G,p_\lambda(\cdot,\cdot))$,
the {\em relative centrality} of a set of nodes $S_1$ with respect to another set of nodes $S_2$, denoted by $C(S_1|S_2)$, is defined as the conditional probability
that  the randomly selected point $Y$ is inside $S_1$ given that the  randomly selected point $X$ is inside $S_2$, i.e.,
\beq{relative0000m}
C(S_1| S_2)=\frac{p_\lambda (S_1,S_2)}{p_{\lambda}(S_2)},
\eeq
where
\bear{exp9911}
&&p_\lambda (S_1,S_2)=\sum_{x \in S_1}\sum_{y \in S_2}p_{\lambda}(x,y),\;\mbox{and}\\
&&p_{\lambda}(S_2)=\sum_{x \in S_2}p_{\lambda}(x).
\eear
\edefin
Clearly, if we choose $S_2=\Omega$,
then the relative centrality of a set of points $S_1$ with respect to $\Omega$ is simply the {\em centrality} of the set of points in $S_1$. Based on the notion of relative centrality, a community can be defined as a set of points that are relatively ``closer'' to each other than to the whole set of data points.

\bdefin{community} {\bf (Community strength and communities (clusters))} For a sampled graph $(G,p_\lambda(\cdot,\cdot))$,
the {\em community strength} of
a set of nodes $S \subset \Omega$, denoted by $Str(S)$,  is defined as the difference of the relative centrality of $S$ with respect to itself and its centrality, i.e.,
\beq{community1111}
Str(S)=C(S|S)-C(S).
\eeq
In particular, if a subset of points $S \subset \Omega$ has a nonnegative community strength, i.e., $Str(S) \ge 0$, then it  is  called
a {\em community} or a {\em cluster} (in this paper, we will use community and cluster interchangeably).
\edefin

As in \cite{chang2013relative}, one can also define the modularity for a partition of a network as the average community strength of a randomly selected point.

\bdefin{index}{\bf (Modularity)}
Consider a sampled graph $(G,p_\lambda(\cdot,\cdot))$.
Let ${\cal P}=\{S_k,k=1,2, \ldots, K\}$, be a partition of $\{1,2,\ldots, n\}$, i.e.,
$S_k \cap S_{k^\prime}$ is an empty set for $k \ne k^\prime$ and $\cup_{k=1}^K S_k=\{1,2,\ldots, n\}$.
The  modularity $Q({\cal P})$ with respect to the partition $S_k$, $k=1,2, \ldots, K$, is defined as the weighted average of the community strength of each subset with the weight being  the centrality of each subset, i.e.,
\beq{index1111}
Q({\cal P})=\sum_{k=1}^K C(S_k)\cdot Str(S_k).
\eeq
\edefin

As the modularity for a partition of $\Omega$ is the average community strength of a randomly selected node, a good partition of a network should have a large modularity. In view of this, one can then tackle the community detection/clustering problem by looking for algorithms that yield large modularity. For this, let us define the covariance between two points $x$ and $y$ in \rdef{covariance} and this will lead to another representation of the modularity.

\bdefin{covariance}{\bf (Covariance)}
For a sampled graph $(G,p_\lambda(\cdot,\cdot))$, the covariance between two points $x$ and $y$ is defined as follows:
\beq{exp8888}
\gsim_\lambda (x,y)=p_\lambda (x,y)-p_{\lambda}(x) p_{\lambda}(y),
\eeq
where $p_{\lambda}(x,y)$ is in \req{exp5555}.
Moreover,  the covariance between two sets $S_1$ and $S_2$ is defined as follows:
\beq{exp9999}
\gsim_\lambda (S_1,S_2)=\sum_{x \in S_1}\sum_{y \in S_2}\gsim_\lambda (x,y).
\eeq
Two sets $S_1$ and $S_2$ are said to be positively correlated if $\gsim_\lambda (S_1,S_2)\ge 0$.
\edefin

It is straightforward to see that
\beq{exp9900}
\gsim_\lambda (S_1,S_2)=p_\lambda (S_1,S_2)-p_{\lambda}(S_1) p_{\lambda}(S_2),
\eeq
and
\bear{index1111b}
Q({\cal P})&=&\sum_{k=1}^K p_\lambda (S_k,S_k)-p_{\lambda}(S_k) p_{\lambda}(S_k)
\nonumber \\
&=&\sum_{k=1}^K \gsim_\lambda(S_k,S_k)
\eear

If we let $\lambda \to 0$, then it is easy to see that
$\gsim_0(x,y)$ is proportional to the semi-cohesion measure $\gsim(x,y)$ in \req{csim7777}.
Thus, the covariance in \req{exp8888} is a generalization of the semi-cohesion measure $\gsim(x,y)$ and it allows us to detect clusters with various resolutions in terms of the average distance $\bar d$. In the next section, we will propose a softmax clustering algorithm that finds a partition to achieve a local maximum of the modularity in \req{index1111b}.

\subsection{The softmax clustering algorithm}

In this section, we propose a probabilistic clustering algorithm, called the {\em softmax clustering algorithm} in Algorithm \ref{alg:Softmax}, based on the softmax function \cite{SoftmaxFunc}. The softmax function  maps a $K$-dimensional vector of arbitrary real values to a $K$-dimensional probability vector. The algorithm starts from a non-uniform probability mass function  for the assignment of each data point to the $K$ clusters. Specifically, let $p_i(k)$  denote the probability that node $i$ is in cluster $k$. Then we repeatedly feed each point to the algorithm to learn the probabilities $p_i(k)'s$. When point $i$ is presented to the algorithm, its expected covariance $z_{i,k}$ to cluster $k$ is computed for $k=1,2,\ldots, K$. Instead of assigning point $i$ to the cluster with the largest positive covariance (the simple maximum assignment in the literature), Algorithm 1 uses a softmax function to update $p_i(k)'s$. Such a softmax update increases (resp. decreases) the confidence of the assignment of point $i$ to clusters with positive (resp. negative) covariances. The softmax update depends on the inverse temperature $\theta$ that is increased every iteration by an annealing parameter $\epsilon$. When $\theta \to\infty$, the softmax update simply becomes the maximum assignment. The ``training'' process is repeated until the objective value $\sum_{k=1}^K\gamma(S_k, S_k)$ converges to a local optimum. The algorithm  then outputs its final partition and the corresponding embedding vector for each data point from  the average covariances  to the $K$ clusters.

\begin{algorithm}[t]
\KwIn{A symmetric  matrix $\Gamma=(\gamma_{ij})$, the number clusters $K$, the inverse temperature $\theta>0$, and the annealing parameter $\epsilon\geq0$.
}
\KwOut{A probabilistic partition of data points  $\{p_i(k),i=1,2,\ldots,n,\;k=1,2,\ldots, K\}$ and an embedding of data points $\{z_i(k),i=1,2,\ldots,n,\;k=1,2,\ldots, K\}$.
}

\noindent {\bf (1)} Set $\gamma_{i, i}=0$ for all $i$.

\noindent {\bf (2)} Initially, each node $i$ is assigned with a (non-uniform) probability mass function $p_i(k)$, $k=1, 2, \ldots, K$ that denotes the probability for node $i$ to be in cluster $k$.;

\noindent {\bf (3)} For $i=1, 2, \ldots, n$

\noindent {\bf (4)} For $k=1, 2, \ldots, K$

\noindent {\bf (5)} Compute $z_i(k)=\sum_{j\neq i}\gamma_{j, i}p_j(k)$.

\noindent {\bf (6)} Let $\tilde p_i(k)=e^{\theta z_i(k)}p_i(k)$, and $c=\frac{1}{\sum_{\ell=1}^K \tilde p_i(\ell)}$.

\noindent {\bf (7)} Update $p_i(k) \Leftarrow c \cdot {\tilde p_i(k)}$
 and $\theta \Leftarrow \theta+\epsilon$.

\noindent {\bf (8)} Repeat from Step 3 until there is no further change.

\caption{The Softmax Clustering Algorithm}
\label{alg:Softmax}
\end{algorithm}

Now we show that Algorithm \ref{alg:Softmax} converges to a local maximum of the objective function $\sum_{k=1}^K\gamma(S_k, S_k)$. For this, we need the following properties in \rlem{StrictlyIncreasing} to show that the objective function is increasing after each update and thus converges to a local optimum in \rthe{ObjIncreasing}.

\blem{StrictlyIncreasing}
Suppose $X$ is a random variable  with the probability mass function $P(X=k)=p_k$, $k=1, 2, \ldots, K$,
and that  $g(X)$ is not a constant (w.p.1) for some $g: \Re\mapsto\Re$. Let
\begin{equation}
\Lambda(\theta)=\log\big(\mathbf{E}[e^{\theta g(X)}]\big).
\end{equation}
Then $\Lambda(\theta)$ is strictly convex and thus $\Lambda^{'}(\theta)$ is strictly increasing in $\theta$.
\elem

\bproof
Since $\Lambda(\theta)=\log\big(\mathbf{E}[e^{g(X)}]\big)$ is  differentiable, we have from the Equation (3.17) that
\begin{equation}
\begin{aligned}
\Lambda{'}(\theta) &= \frac{\mathbf{E}[g(X)e^{\theta g(X)}]}{\mathbf{E}[e^{\theta g(X)}]}, \\
\Lambda{''}(\theta) &= \frac{\big(\mathbf{E}[(g(X))^2 e^{\theta g(X)}]\big)\big(\mathbf{E}[e^{\theta g(X)}]\big) - \big(\mathbf{E}[g(X)e^{\theta g(X)}]\big)^2}{\big(\mathbf{E}[e^{\theta g(X)}]\big)^2}.
\end{aligned}
\end{equation}
From the Schwartz inequality, we know that for any two random variables $Z_1$ and $Z_2$.
\beq{cauchy1111}
|\mathbf{E}[Z_1 Z_2]|^2\leq(\mathbf{E}[Z_1^2])(\mathbf{E}[Z_2^2]).
\eeq
Moreover, the above inequality becomes an equality only when $Z_1 =cZ_2$ (w.p.1) for some constant $c$.
From \req{cauchy1111} and $\theta\geq0$, it follows that
\[
\mathbf{E}\big[(g(X)e^{\frac{\theta}{2}g(X)})^2\big] \mathbf{E}\big[(e^{\frac{\theta}{2}g(X)})^2\big] - \big(\mathbf{E}\big[g(X)e^{\frac{\theta}{2}g(X)}\cdot g(X)e^{\frac{\theta}{2}g(X)}\big]\big)^2\geq0
\]
As $g(X)$ is not a constant (w.p.1), the above inequality is strict.
Thus, we have
\[
\Rightarrow \Lambda^{''}(\theta)>0.
\]
\eproof

Note from \rlem{StrictlyIncreasing} that $\Lambda^{'}(0)=\mathbf{E}[g(X)]$ and thus $\Lambda^{'}(\theta)>\Lambda^{'}(0)$ for $\theta>0$.  This leads to the following corollary.

\bcor{inccor} Under the assumptions in \rlem{StrictlyIncreasing}, for all $\theta > 0$,
$$\frac{\mathbf{E}[g(X)e^{\theta g(X)}]}{\mathbf{E}[e^{\theta g(X)}]} > \mathbf{E}[g(X)].$$
\ecor

For more mathematically rigorous derivations, see \cite{ProfSoftmax}, Lemma 8.1.5, Corollary 8.1.6, Proposition 7.1.4, and Proposition 7.1.8.

\bthe{ObjIncreasing}
Given a symmetric matrix $\Gamma=(\gamma_{ij})$ with $\gamma_{ii}=0$ for all $i$, the following objective value
\beq{objc1111}
 \sum\limits_{k=1}^K\sum\limits_{i=1}^n\sum\limits_{j=1}^n\gamma_{ij}p_i(k)p_j(k)
\eeq
is increasing after each update in Algorithm \ref{alg:Softmax}.
Thus, the the objective values converge monotonically to a finite constant.
\ethe

\bproof
Suppose that $i_0$ is the point updated in Step (6) of Algorithm \ref{alg:Softmax}.
After the update, the objective value in \req{objc1111} can be written as follows:
\beq{objc2222}
\sum\limits_{k=1}^K\bigg(\sum\limits_{i\neq i_0}\sum\limits_{j\neq i_0}\gamma_{ij}p_i(k)p_j(k) + \sum\limits_{j\neq i_0}\gamma_{i_0 j}\tilde{p}_{i_0}(k)p_j(k) + \sum\limits_{i\neq i_0}\gamma_{i i_0}{p}_i(k)\tilde{p}_{i_0}(k) + \gamma_{i_0}\gamma_{i_0}\tilde{p}_{i_0}(k)\tilde{p}_{i_0}(k)\bigg).
\eeq
Since $\gamma_{ij}=\gamma_{ji}$ (symmetric) and $\gamma_{ii}=0$ for all $i$,  we have from \req{objc2222} that the difference, denoted by $\Delta_{Obj}$, between the objective value {\em after} the update and that {\em before} the update can be computed as follows:
\begin{equation}
\begin{aligned}
\Delta_{Obj} &= \sum\limits_{k=1}^K 2\cdot\bigg[\sum\limits_{j\neq i_0}\gamma_{i_0 j}\tilde{p}_{i_0}(k)p_j(k) - \sum\limits_{j\neq i_0}\gamma_{i_0 j}p_{i_0}(k)p_j(k)\bigg] \\
&= 2\cdot\sum\limits_{k=1}^K\bigg[\tilde{p}_{i_0}(k)\bigg(\sum\limits_{j\neq i_0}\gamma_{i_0 j}p_j(k)\bigg) - p_{i_0}(k)\bigg(\sum\limits_{j\neq i_0}\gamma_{i_0 j}p_j(k)\bigg)\bigg] \\
&= 2\cdot\sum\limits_{k=1}^K\bigg[\tilde{p}_{i_0}(k)z_{i_0}(k) - p_{i_0}(k)z_{i_0}(k)\bigg].
\end{aligned}
\end{equation}
Now view $z_i(k)$ as $g(k)$, $p_{i,0}(k)$ as $P(X=k)$ in \rlem{StrictlyIncreasing}.
Then
\beq{objc3333}
\tilde{p}_{i_0}(k) = \frac{e^{\theta z_i(k)}p_i(k)}{\sum\limits_{\ell=1}^K e^{\theta z_i(\ell)}p_i(\ell)} = \frac{e^{\theta g(k)}p_{i_0}(k)}{\mathbf{E}[e^{\theta g(X)}]}.
\eeq
Note from \req{objc3333} that $\Delta_{Obj}$ can also be written as follows:
\begin{equation}
\begin{aligned}
\Delta_{Obj} &= 2\cdot\bigg[\sum\limits_{k=1}^K z_{i_0}(k)\frac{e^{\theta z_{i_0}(k)}p_{i_0}(k)}{\sum\limits_{\ell=1}^K e^{\theta z_{i_0}(\ell)}p_{i_0}(\ell)} - \sum\limits_{k=1}^K p_{i_0}(k)z_{i_0}(k)\bigg] \\
&= 2\bigg[\frac{\mathbf{E}[g(X)e^{\theta g(X)}]}{\mathbf{E}[e^{\theta g(X)}]} - \mathbf{E}[g(X)]\bigg].
\end{aligned}
\end{equation}
From the \rcor{inccor}, we conclude that $\Delta_{Obj}>0$ for all $\theta>0$. Since the objective values in \req{objc1111} are bounded, the objective values converge monotonically to a finite constant.
\eproof

Now we explain the reason why we have to initialize each node with a non-uniform probability mass function $p_i(k)$. Suppose the probability mass function $p_i(k)$ is a uniform distribution, i.e., $p_i(k)=\frac{1}{K}$ for all $i$ and $k$. We then have
\begin{equation}
z_i(k) = \sum\limits_{j\neq i}\gamma_{ji}p_j(k) = \bigg(\sum\limits_{j\neq i}\gamma_{ji}\bigg)\frac{1}{K}.
\end{equation}
Thus,
\begin{equation}
z_i(1) = z_i(2) = \ldots = z_i(K) = \beta,
\end{equation}
for some constant $\beta$.
This then leads to
\begin{equation}
\tilde{p}_i(k) = \frac{e^{\theta\beta}\cdot\frac{1}{K}}{\sum\limits_{\ell=1}^K e^{\theta\beta}\cdot\frac{1}{K}} = \frac{1}{K}.
\end{equation}
Thus, the probability mass function $p_i(k)$ will remain the same as the initial assignment after each update, and the algorithm is trapped in a bad local optimum. On the contrary, if we start from a non-uniform probability mass function for each node, then it is less likely to be trapped in a bad local optimum.

Recall that in each update
\begin{equation}
\tilde{p}_i(k) = \frac{e^{\theta z_i(k)}p_i(k)}{\sum\limits_{\ell=1}^K e^{\theta z_i(\ell)}p_i(\ell)},\quad k = 1, 2, \ldots, K.
\end{equation}
When $\theta$ is very large, the denominator of $\tilde{p}_i(k)$ can be approximated as follows:
\begin{equation}
\sum\limits_{\ell=1}^K e^{\theta z_i(\ell)}p_i(\ell)\approx\max_{1\leq \ell\leq K}e^{\theta z_i(\ell)}p_i(\ell).
\end{equation}
Let $k_0 = \arg\max_{1\leq \ell\leq K}e^{\theta z_i(\ell)}p_i(\ell)$.
Then
\begin{equation}
\tilde{p}_i(k) \xrightarrow{\theta\rightarrow\infty} \delta_{i, k_0}=
\begin{cases}
1, &i=k_0 \\
0, &i\neq k_0
\end{cases},
\end{equation}
which shows that the softmax update becomes the maximum assignment when $\theta \to \infty$. As the inverse temperature $\theta$ is increased by  the annealing parameter $\epsilon$ after each update, we conclude that eventually the softmax update becomes the maximum assignment in Algorithm \ref{alg:Softmax}. Thus, Algorithm \ref{alg:Softmax} will output a deterministic partition when the algorithm converges.


\subsection{An illustrating experiment with three rings}

\begin{figure}[htbp]
	\centering
	\includegraphics[width=0.70\textwidth]{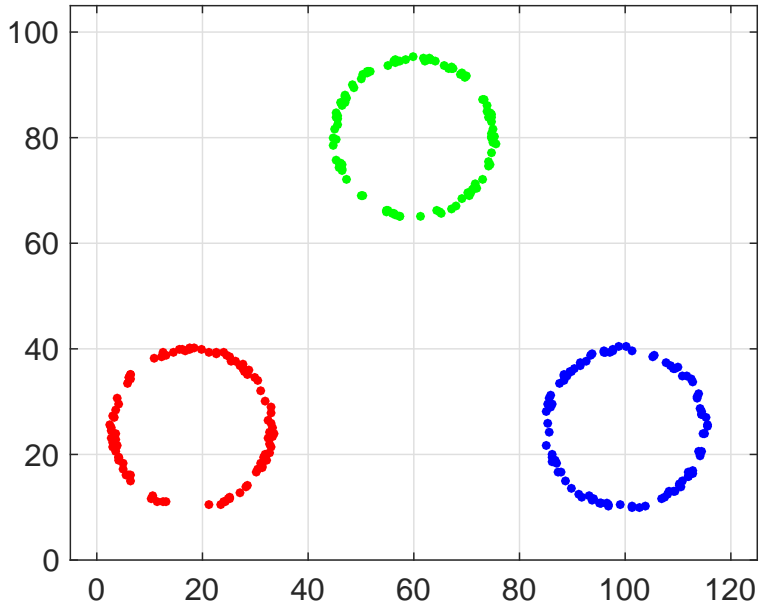}
	\caption{The ground-truth plot of the three-ring dataset}
	\label{fig:OPD_3Ring}
\end{figure}

\begin{figure*}[htbp]
    \centering
    \begin{tabular}{p{0.35\textwidth}p{0.1\textwidth}p{0.35\textwidth}}
	\centering
    	\includegraphics[width=0.35\textwidth]{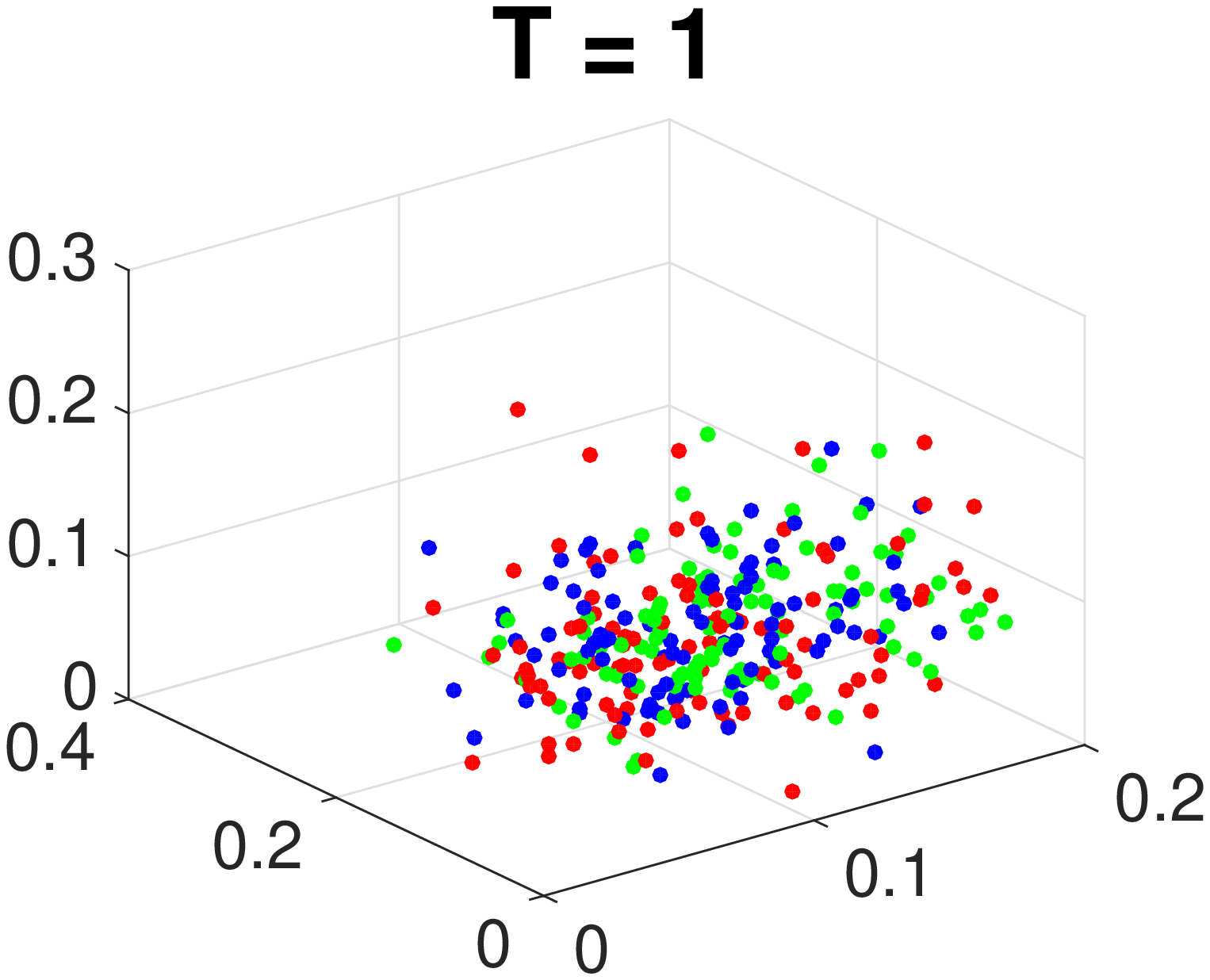} & &
    	\includegraphics[width=0.35\textwidth]{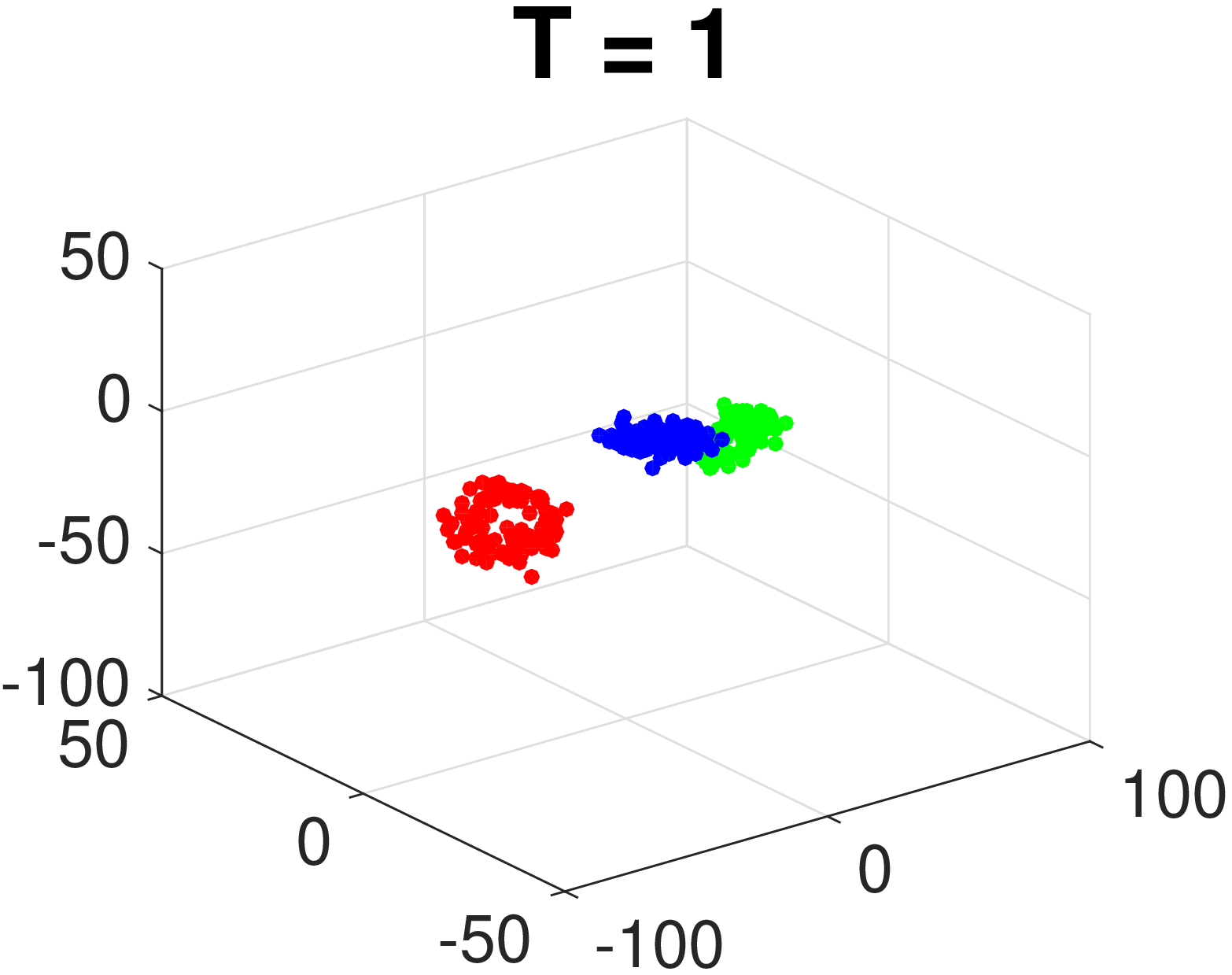} \\
    \centering
    	\includegraphics[width=0.35\textwidth]{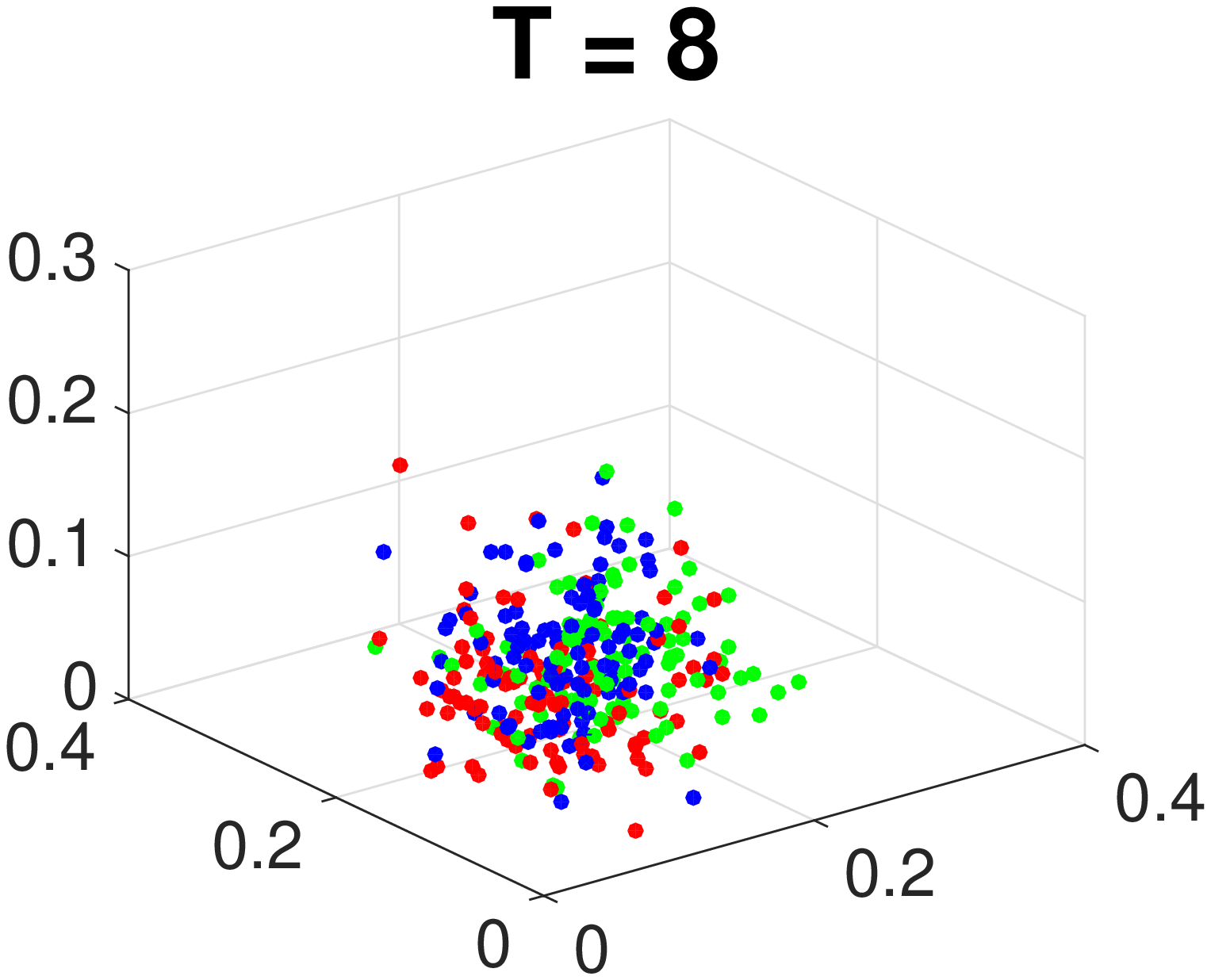} & &
      	\includegraphics[width=0.35\textwidth]{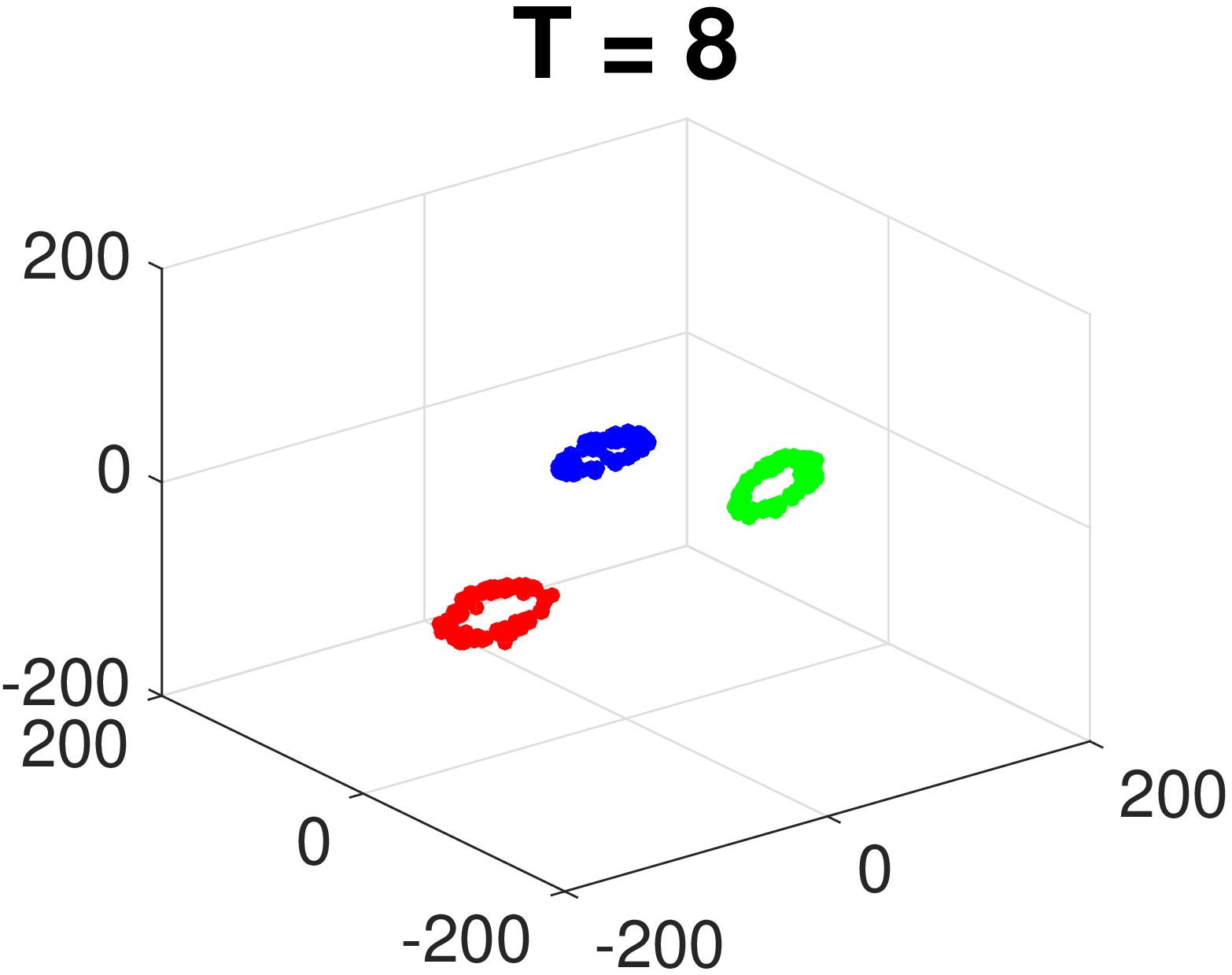} \\
    \centering
      	\includegraphics[width=0.35\textwidth]{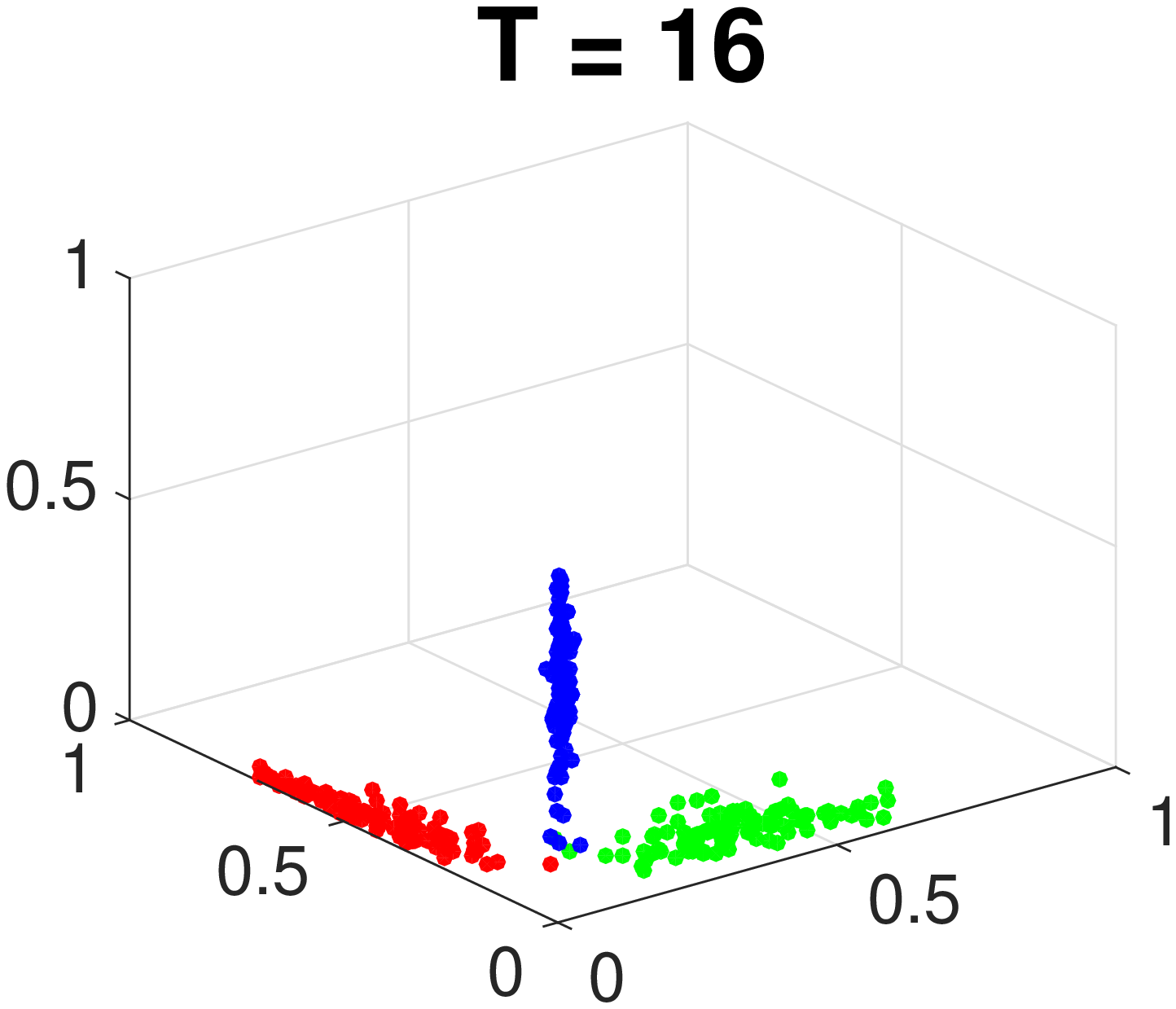} & &
      	\includegraphics[width=0.35\textwidth]{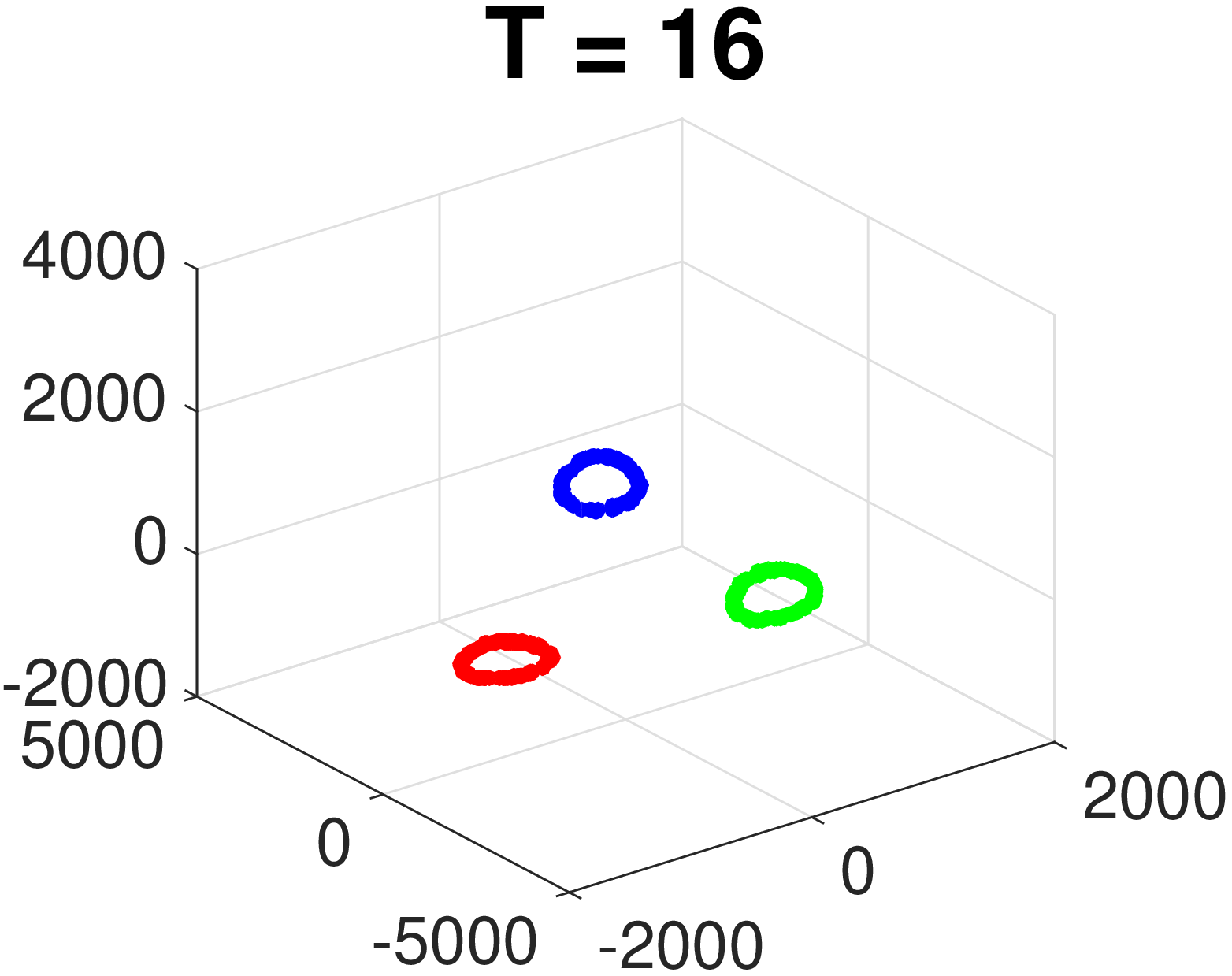} \\
    \centering
      	\includegraphics[width=0.35\textwidth]{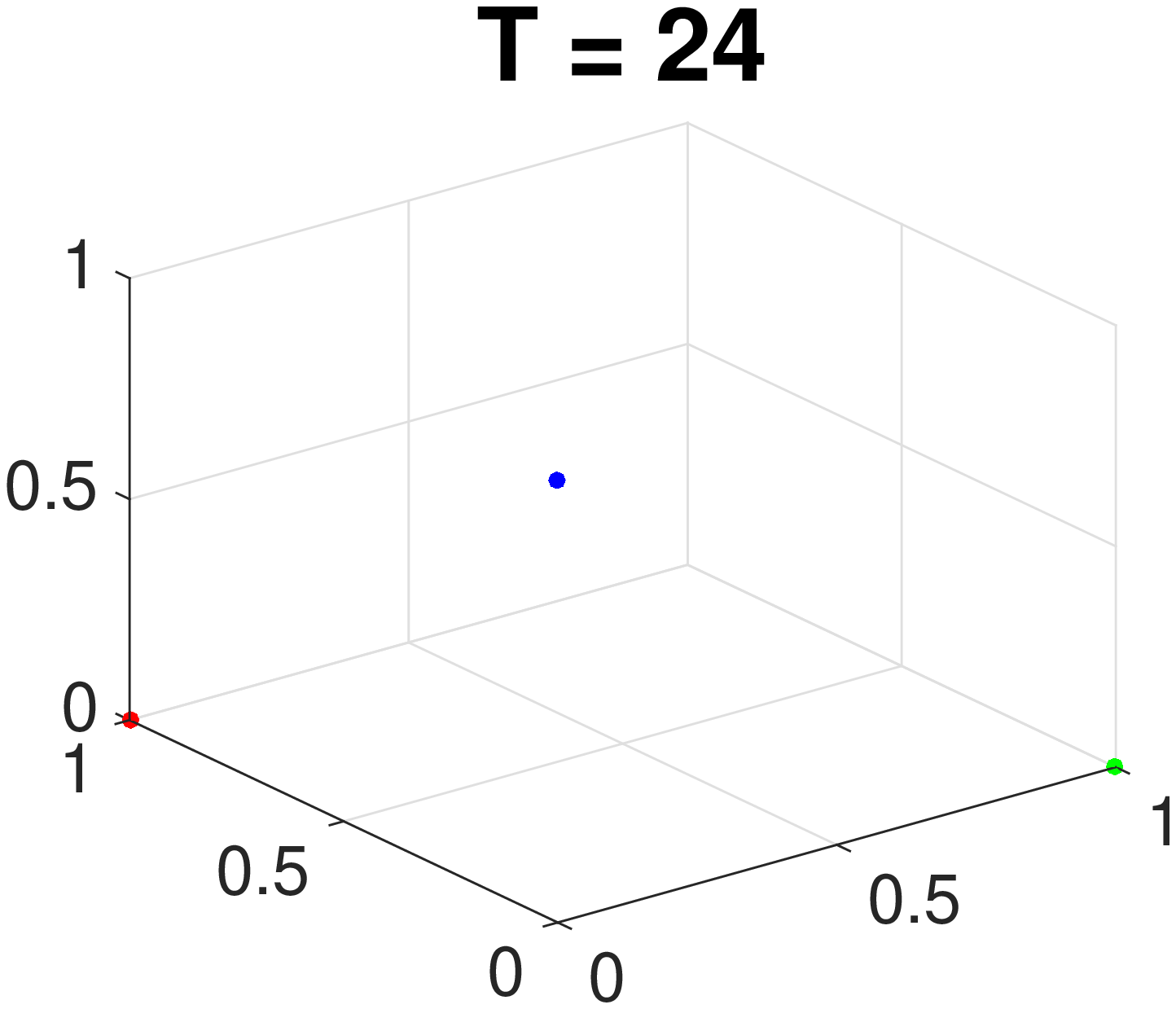} & &
      	\includegraphics[width=0.35\textwidth]{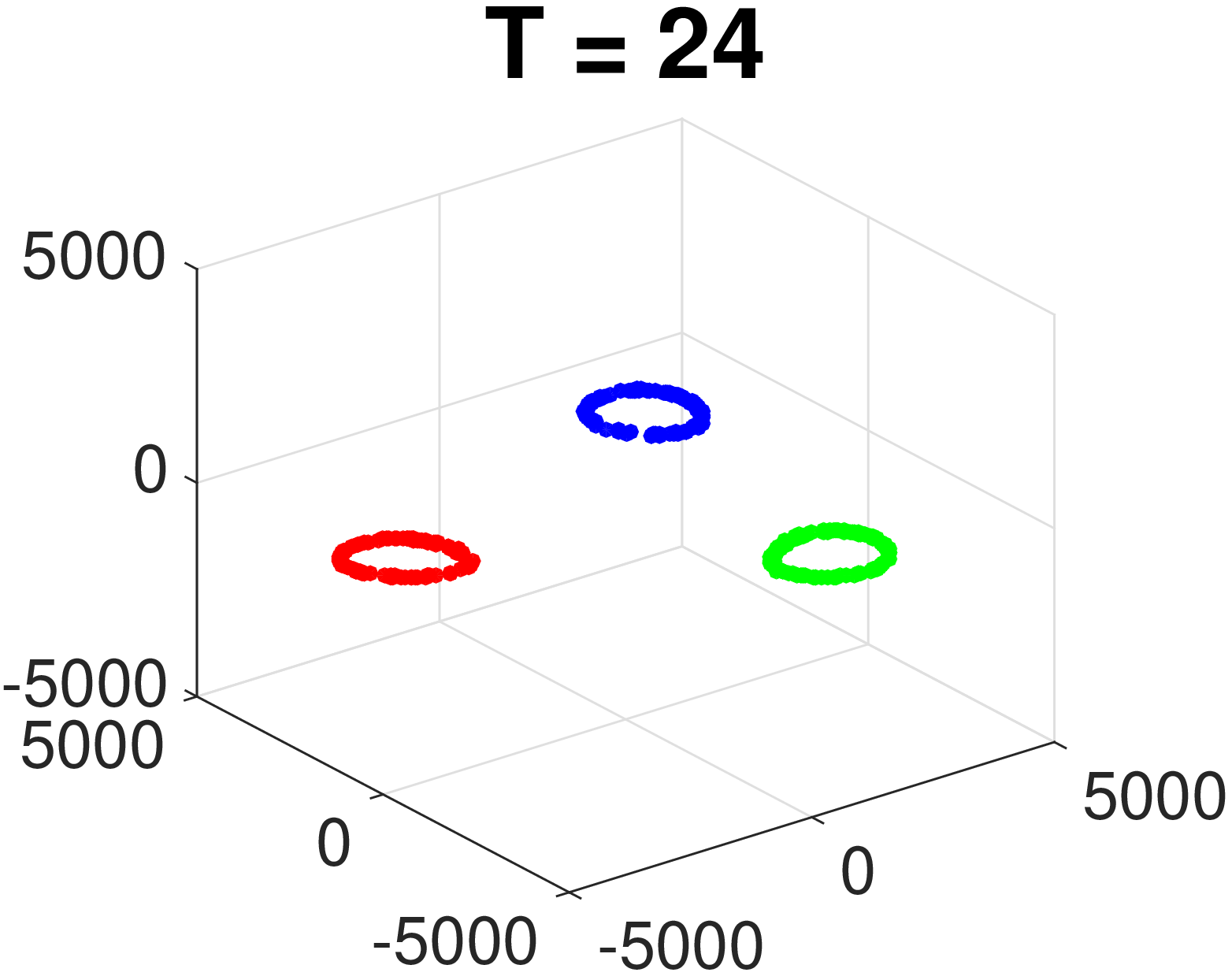} \\
    \centering
      	(a) $p_i(k)$'s in a 3-D plot& &
    \centering
      	(b) $z_i(k)$'s in a 3-D plot
    \end{tabular}
    \caption{$p_i(k)$'s and $z_i(k)$'s  for $K=6$, $k \in \{3, 5, 6\}$ in 3-D plots for various training iterations $T$.}
    \label{fig:ResultSCalg}
\end{figure*}

In this section, we conduct an illustrating experiment for the softmax clustering algorithm. For this experiment, we generate an artificial dataset with three non-overlapping rings (as shown in \rfig{OPD_3Ring}). The total number of nodes $n$ is $300$ with $100$ nodes in each ring (cluster). Though the plot in \rfig{OPD_3Ring} reveals that these 300 points are in fact in a two-dimensional space, one can image that these 300 points might be given from a very high dimensional Euclidean space (this can be easily done by using a measure preserving transformation that transform theses 300 points into a very high dimensional Euclidean space). For this experiment, we only need the Euclidean distance between any two points and we suppose that we are given a $300 \times 300$ distance matrix (that might be in fact from a  very high dimensional Euclidean space). We then use \req{csim7777} to compute the semi-cohesion measure (matrix) and use that matrix as the input of Algorithm \ref{alg:Softmax}. For the other inputs of Algorithm \ref{alg:Softmax}, we set the number of clusters $K=6$, the inverse temperature $\theta=0.00025$ and the annealing parameter $\epsilon=0.000025$. Though the number of clusters is set to be 6 at the beginning, the softmax clustering algorithm converges to only three clusters (that match the number of ground-truth clusters). Even though both the probabilistic partition $(p_i(1),p_i(2),\ldots, p_i(6))$ and the corresponding embedding vector $(z_i(1),z_i(2), \ldots, z_i(6))$ of  point $i$ are the $6$-dimensional  vectors, there are only three nonzero coordinates in the probabilistic partition vectors when the algorithm converges. In \rfig{ResultSCalg}, we show our experimental results for the probabilistic partition $p_i(k)'s$ and the corresponding embedding vector $z_i(k)'s$ in 3D plots by using the final three nonzero coordinates, i.e., $k \in \{3,5,6\}$. Each node is marked with the same color as that in its ground-truth ring in \rfig{OPD_3Ring}. The parameter $T$ is the number of iterations that the $n$ data points are used for ``training'' the algorithm. In particular, for $T=1$, the probabilistic partition $p_i(k)'s$ is a random vector as we do not know how to classify a point to a cluster yet. As $T$ increases,  one can see that the softmax clustering algorithm is gaining some ``confidence'' on how a point should be classified to these $K$ clusters after several iterations of ``training.'' At $T=15$, each probabilistic partition vector converges to a delta function and that results in a deterministic partition of three clusters for the $n$ points. On the other hand, we can also see that the ``topology'' of the three rings start to emerge from our embedding vectors as $T$ increases. Interestingly enough, if we knew the final three coordinates (before any training), then we can see from the plot of $z_i(k)$ at  $T=1$ in Figure \ref{fig:ResultSCalg} (b) that the data points have been already divided into three clusters. From this illustrating experiment, we demonstrate that our softmax clustering algorithm not only  performs well for clustering  but also allows us to visualize the original manifold of the input data by embedding them into a low-dimensional Euclidean space. Finally, we note that it is better that we set the initial number of clusters $K$  to be larger than the number of ground-truth clusters as otherwise we might force the algorithm to merge several ground-truth clusters into a single cluster. It is also not necessary to start from $K=n$ as it might be costly to compute.

\subsection{Further supporting evidence by using the eigendecomposition of the semi-cohesion measure}

\begin{figure}[htbp]
	\centering
	\includegraphics[width=0.75\textwidth]{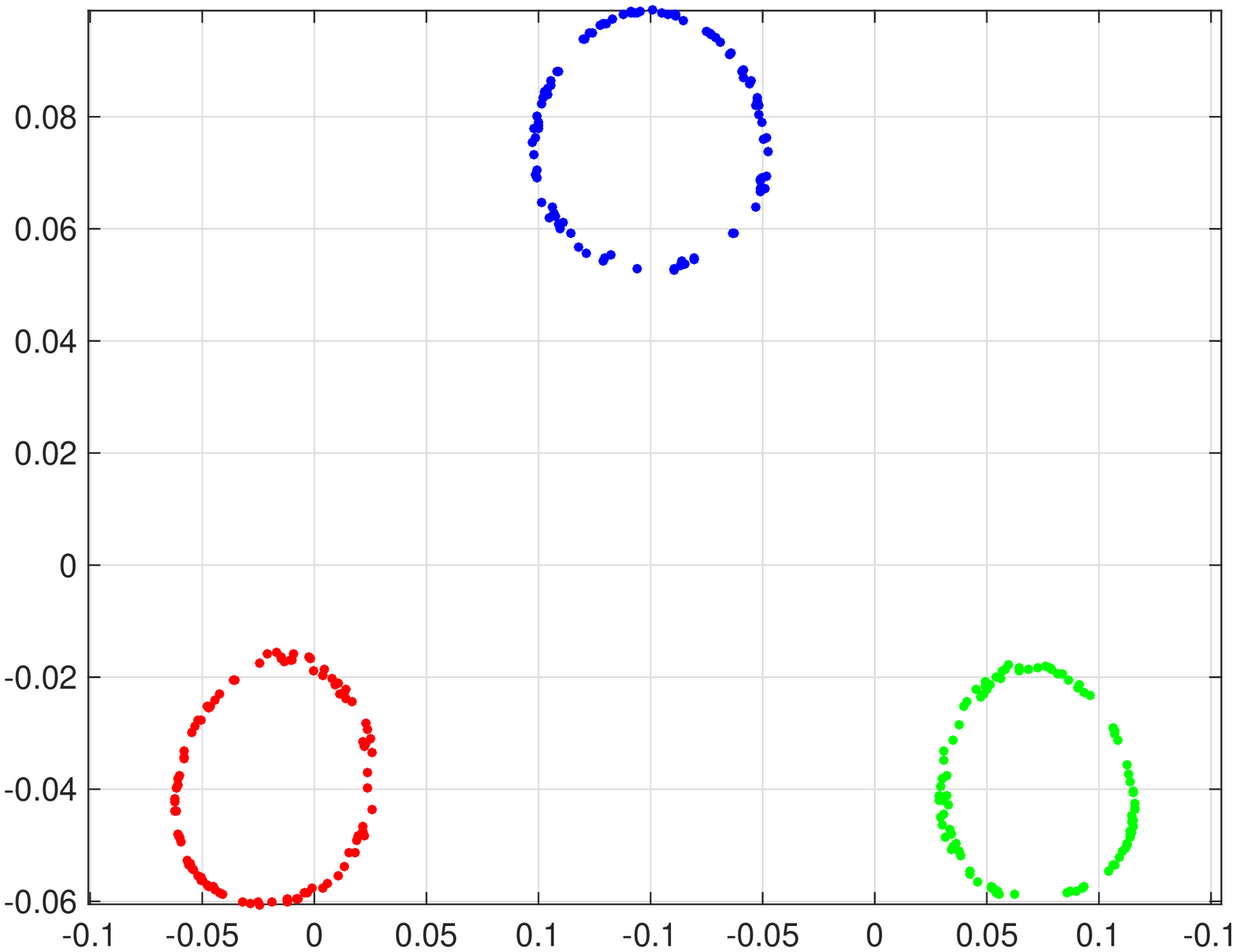}
	\caption{Embedding by using the largest two eigenvectors as the X-Y coordinates of each point for the three-ring dataset.}
	\label{fig:ThreeRingsEig}
\end{figure}

\begin{figure}[htbp]
	\centering
	\includegraphics[width=0.8\textwidth]{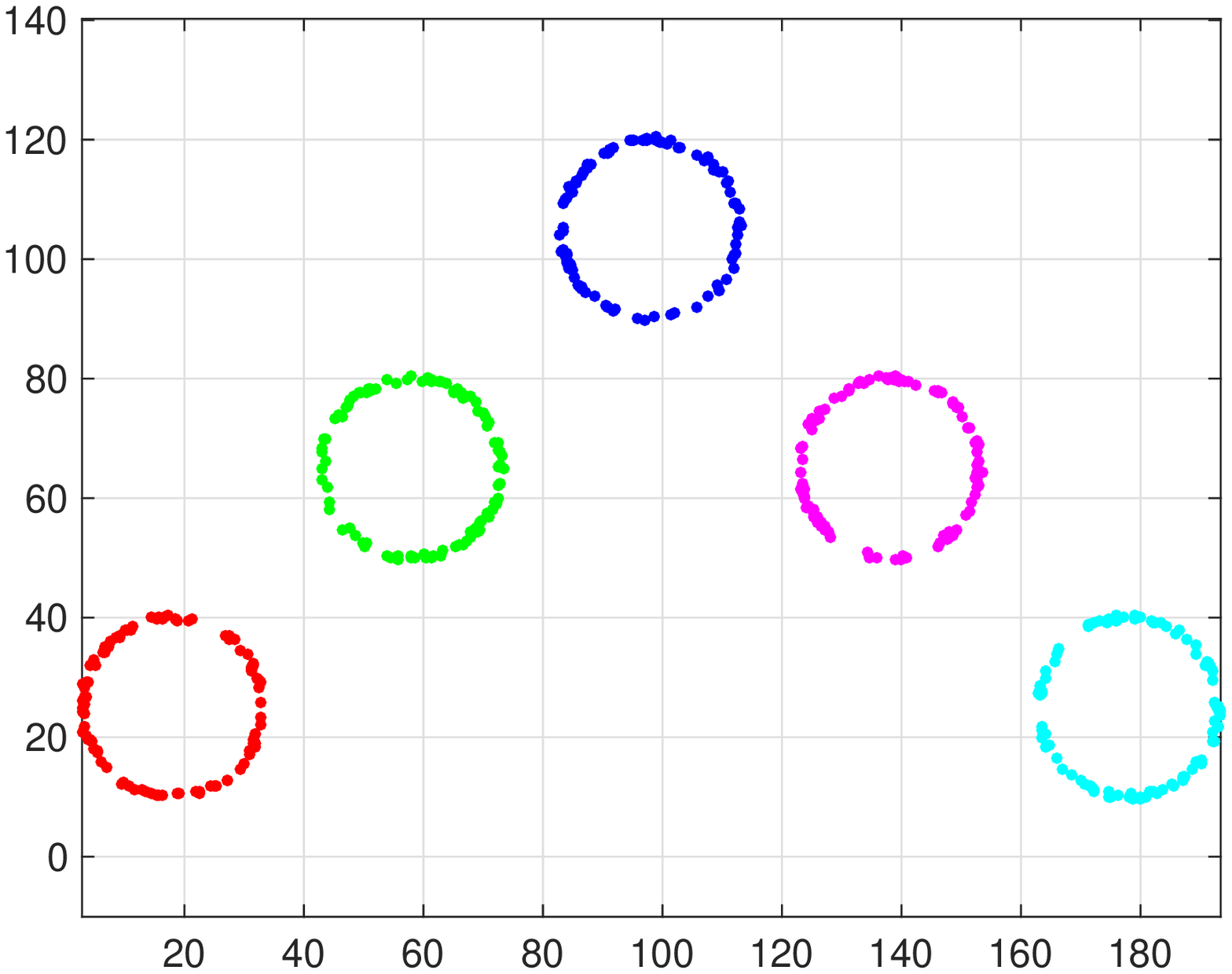}
	\caption{The ground-truth plot of the five-ring dataset.}
	\label{fig:FiveRingsGT}
\end{figure}

\begin{figure}[htbp]
	\centering
	\includegraphics[width=0.75\textwidth]{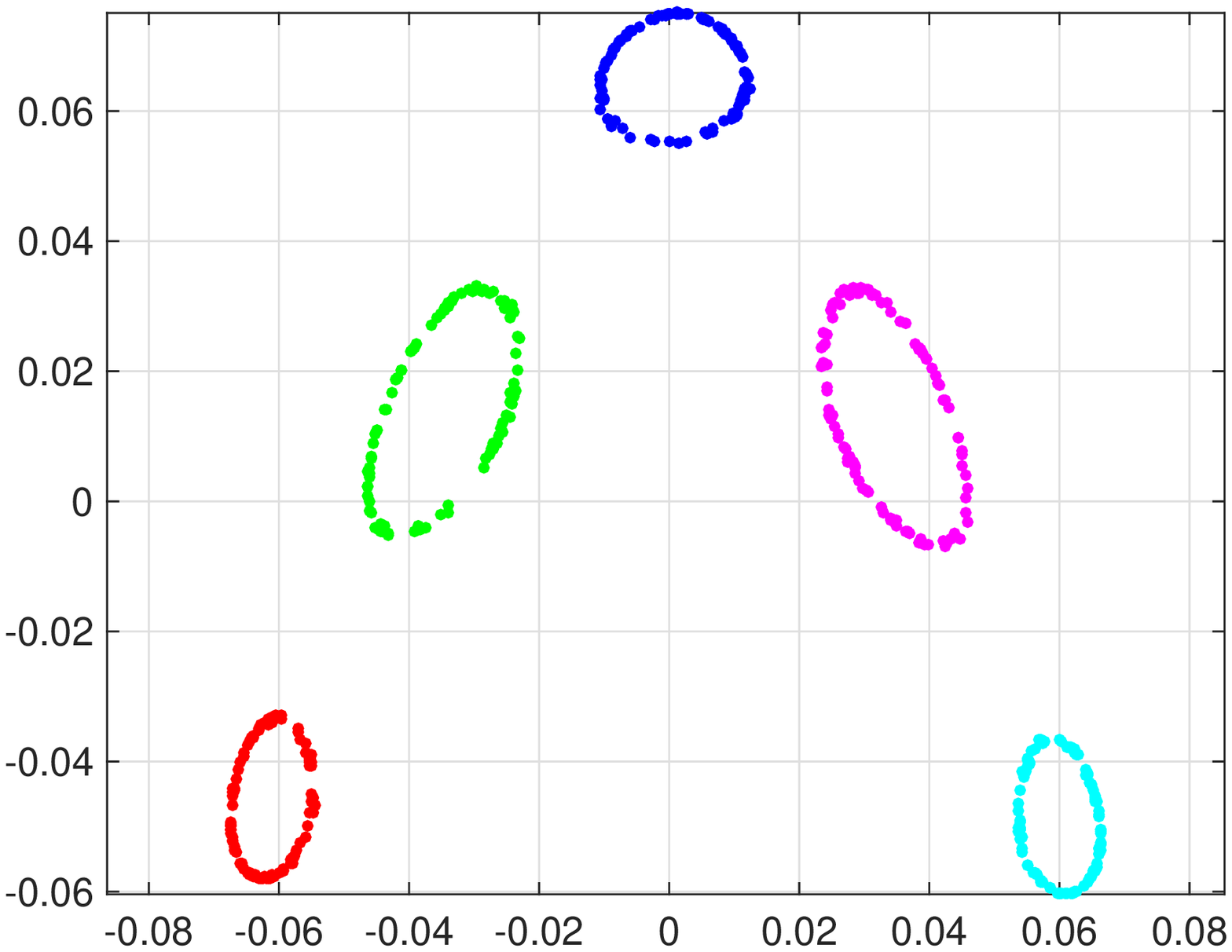}
	\caption{Embedding by using the largest two eigenvectors as the X-Y coordinates of each point for the five-ring dataset.}
	\label{fig:FiveRingsEig}
\end{figure}

In this section, we provide further supporting evidence for our softmax clustering algorithm that uses the semi-cohesion matrix as its input. We do this by using the eigendecomposition of the semi-cohesion measure matrix. In \rfig{ThreeRingsEig}, we use the largest two eigenvalues and their corresponding eigenvectors to represent the x- and y-coordinate of each point in the two-dimensional space. Comparing to the ground-truth plot in \rfig{OPD_3Ring}, one can see that the original ``topology'' of the manifold  can be ``preserved'' by projecting these two eigenvectors on a two-dimensional space. To further versify this, we  construct another artificial dataset with five non-overlapping rings in a two-dimensional space (see \rfig{FiveRingsGT}). The total number of nodes  $n$ is $500$ with 100 nodes in each cluster/ring. As shown in \rfig{FiveRingsGT} and \rfig{FiveRingsEig}, the original ``topology'' of the manifold  can also be ``preserved'' by projecting these two eigenvectors on a two-dimensional space.

\begin{figure}[htbp]
	\centering
	\includegraphics[width=0.55\textwidth]{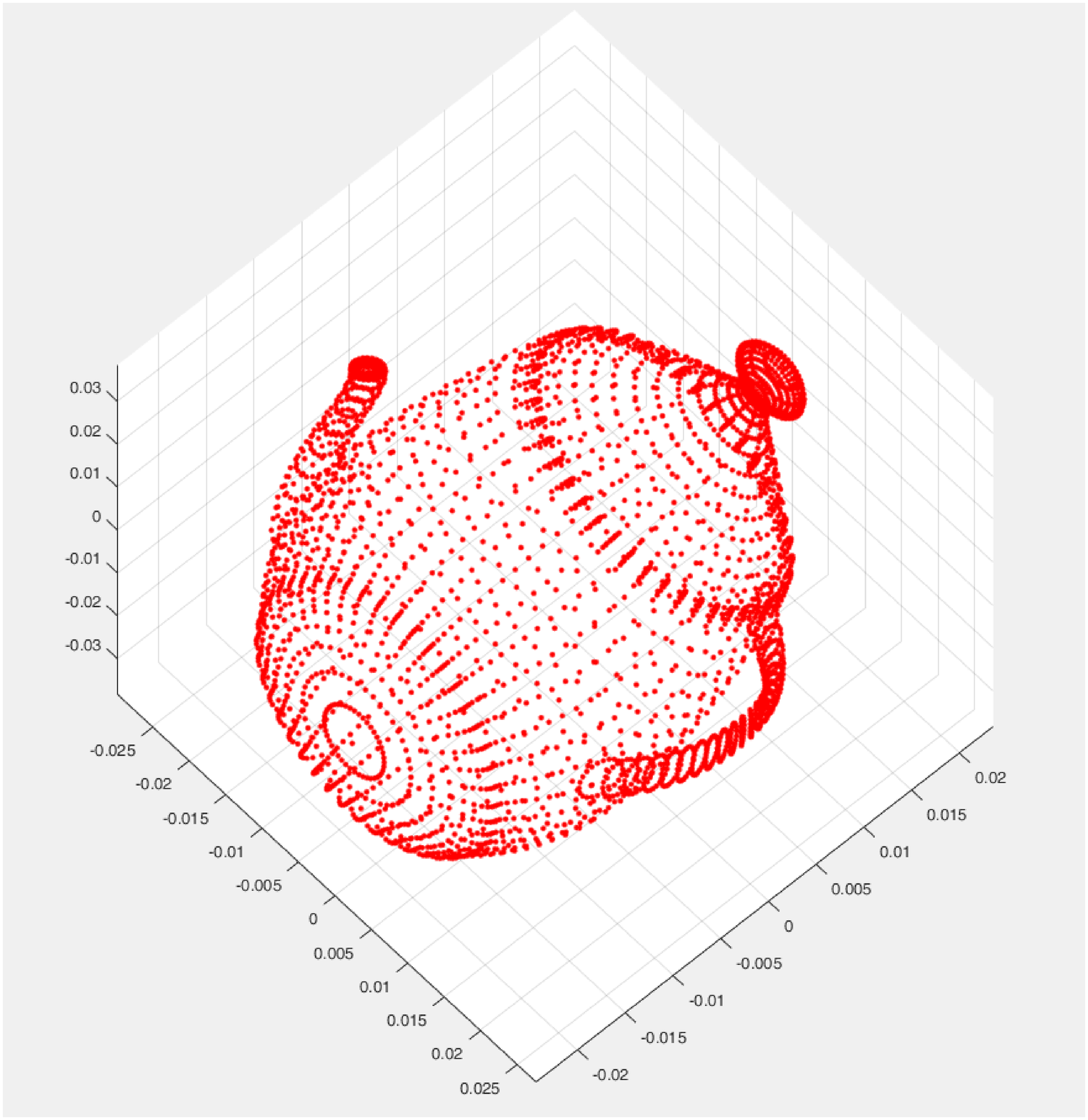}
	\caption{Use the largest three of eigenvectors as X-Y-Z coordinates of each point}
	\label{fig:TeapotEig}
\end{figure}

In the following, we use the MATLAB built-in function called ``teapotGeometry'' to generate a teapot plot in the three-dimensional space. By using the largest three eigenvalues of the semi-cohesion matrix and their corresponding eigenvectors to represent x-, y- and z-coordinate in the three-dimensional space,  we can see from  \rfig{TeapotEig} that the original manifold still can be preserved by projecting these three eigenvectors on a three-dimensional space.

\begin{figure}[htbp]
	\centering
	\includegraphics[width=1\textwidth]{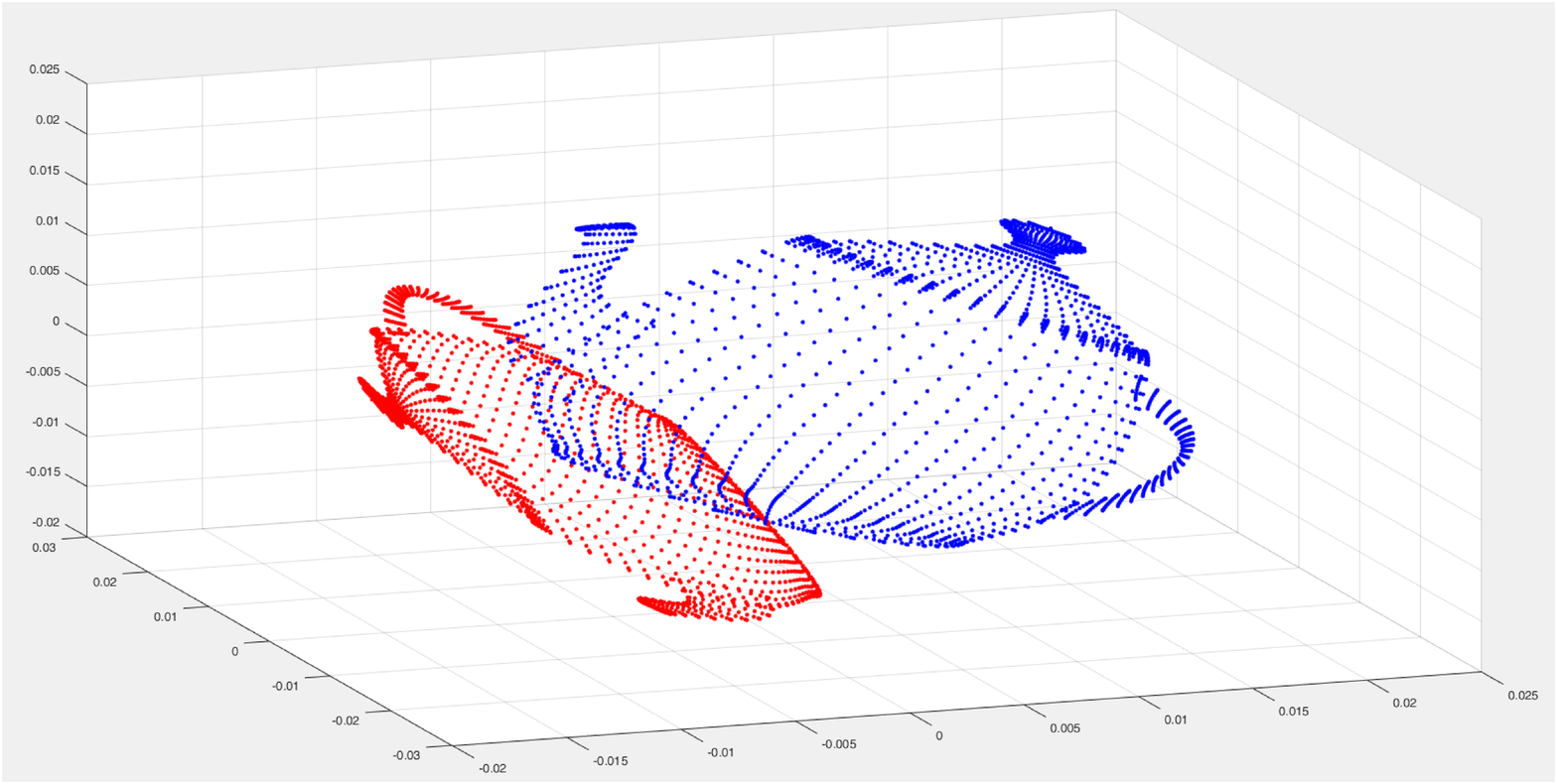}
	\caption{Two teapots from 5D to 3D embedding}
	\label{fig:5D_TeapotEig}
\end{figure}

To test the high-dimensional data, we  generate a five-dimensional dataset that contains two teapots of the same size: one is located in the one-, two-, three-dimensional space, and the other is in the three-, four-, five-dimensional space. As shown in \rfig{5D_TeapotEig}, we still preserve the original manifold even though we are not using the largest three eigenvalues and  their corresponding eigenvectors. However, if we use the corresponding eigenvectors from small eigenvalues, there will be significant interference and that makes the original manifold very difficult to see.

\begin{figure}[htbp]
	\centering
	\includegraphics[width=1\textwidth]{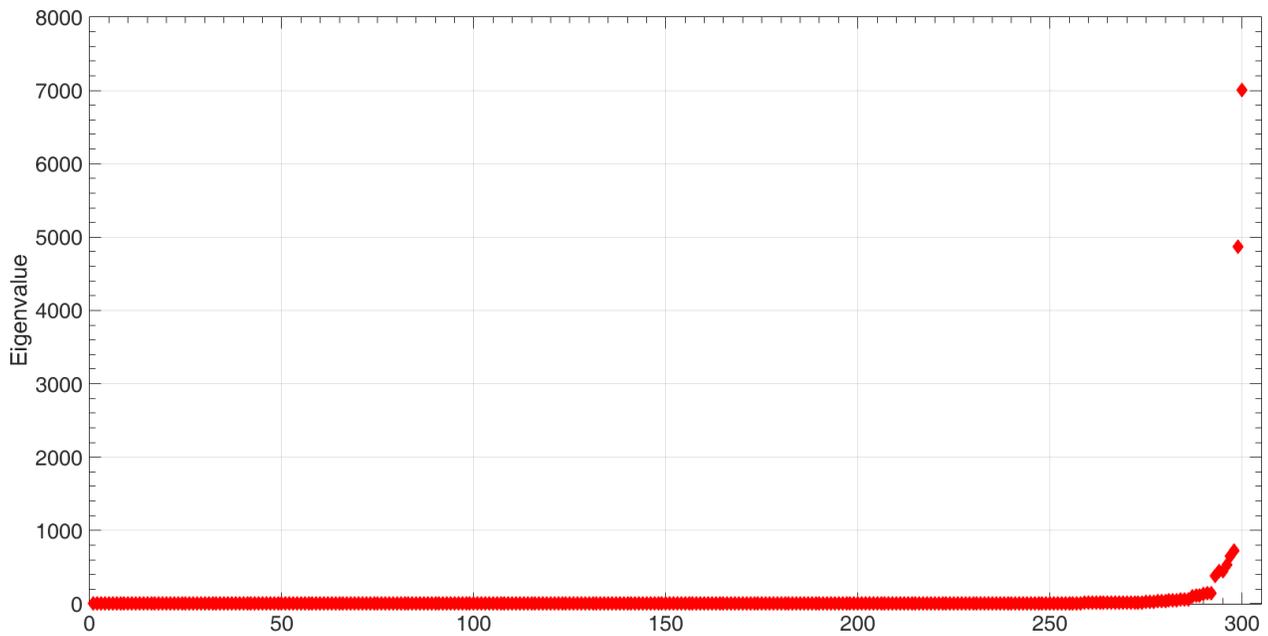}
	\caption{The distribution of eigenvalues for the three-ring dataset.}
	\label{fig:3RingsDist}
\end{figure}

\begin{figure}[htbp]
	\centering
	\includegraphics[width=1\textwidth]{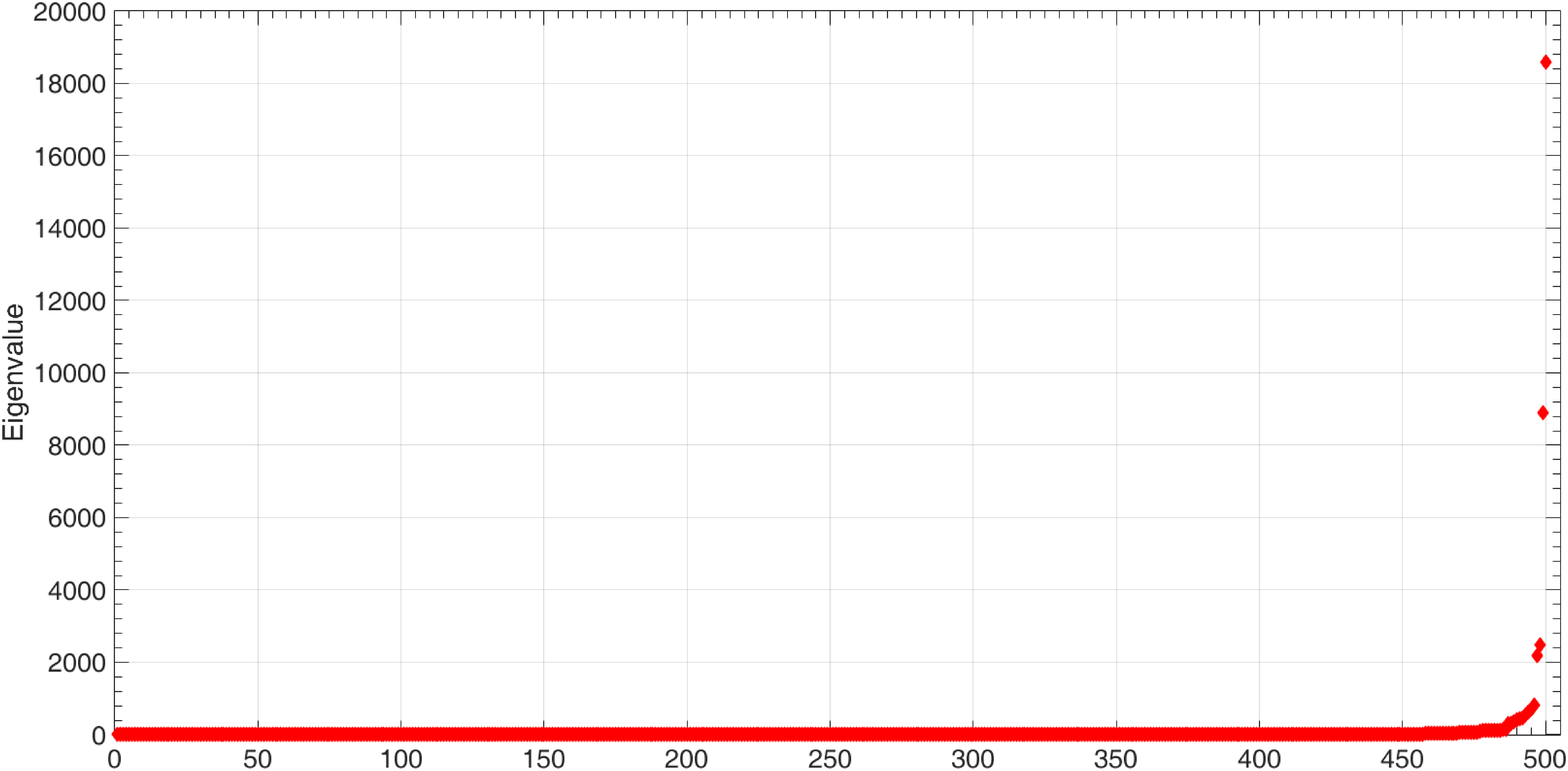}
	\caption{The distribution of eigenvalues for the five-ring dataset.}
	\label{fig:5RingsDist}
\end{figure}

In \rfig{3RingsDist} and \rfig{5RingsDist}, we plot the distributions of the eigenvalues for the two semi-cohesion  matrices obtained from the three-ring dataset in  \rfig{OPD_3Ring} and the five-ring dataset in \rfig{FiveRingsGT}. As shown in \rfig{3RingsDist} and \rfig{5RingsDist}, these two distributions are very similar, and they both have a significant gap between the largest two of eigenvalues and the others. One surprise finding is that all  the eigenvalues of the two semi-cohesion  matrices are non-negative, which implies that the semi-cohesion  matrix obtained from an Euclidean distance in $\mathbb{R}^p$ is a positive semi-definite matrix. We will discuss its connections to PCA further in the next section.

\subsection{Connections to PCA}

In this section, we discuss the connections between the eigendecomposition of a semi-cohesion measure and the  principal component analysis (PCA). As all the data points are in a high dimensional Euclidean space in our previous experiments, let us consider using the {\em squared} Euclidean distance to generate the semi-cohesion measure. For this, let $d_{hs}(x, y)$ be one-half of the squared Euclidean distance between any two points $x$ and $y$ in a high dimensional Euclidean space, i.e.,
\begin{equation}
d_{hs}(x, y)=\frac{1}{2}(x-y)^T(x-y).
\end{equation}
Thus, the semi-cohesion measure $\gamma(\cdot,\cdot)$ in \req{csim7777} can be computed  by
\begin{equation}
\begin{aligned}
\gamma(x, y ) &= \frac{1}{2n^2}\sum_{z_2\in\Omega}\sum_{z_1\in\Omega}\big[(x-z_1)^T(x-z_1)+(y-z_2)^T(y-z_2) \\
& \qquad\qquad\qquad -(z_1-z_2)^T(z_1-z_2)-(x-y)^T(x-y)\big] \\
&= \frac{1}{2n^2}\sum_{z_2\in\Omega}\sum_{z_1\in\Omega}\big[x^Tx-2z_1^Tx+z_1^Tz_1+y^Ty-2z_2^Ty+z_2^Tz_2 \\
& \qquad\qquad\qquad -z_1^Tz_1+2z_1^Tz_2-z_2^Tz_2-x^Tx+2x^Ty-y^Ty\big] \\
&= \frac{1}{n^2}\sum_{z_2\in\Omega}\sum_{z_1\in\Omega}(x^Ty+z_1^Tz_2-z_1^Tx-z_2^Ty) \\
&= \frac{1}{n^2}\sum_{z_2\in\Omega}\sum_{z_1\in\Omega}x^T(y-z_1)-z_2^T(y-z_1) \\
&= \frac{1}{n^2}\sum_{z_1\in\Omega}\big(\sum_{z_2\in\Omega}(x-z_2)^T\big)(y-z_1) \\
&= \big(\frac{1}{n}\sum_{z_2\in\Omega}(x-z_2)^T\big)\big(\frac{1}{n}\sum_{z_1\in\Omega}(y-z_1)\big) \\
&= \big(x-\frac{1}{n}\sum_{z_2\in\Omega}z_2\big)^T\big(y-\frac{1}{n}\sum_{z_1\in\Omega}z_1\big).
\end{aligned}
\label{eq:pca1111}
\end{equation}
Let
\begin{equation}
c = \frac{1}{n}\sum_{z_2\in\Omega}z_2=\frac{1}{n}\sum_{z_1\in\Omega}z_1
\end{equation}
be the centroid of all the points in the dataset.
Then the semi-cohesion measure in \req{pca1111} can be written as follows:
\begin{equation}
\gamma(x, y ) = (x-c)^T(y-c).
\end{equation}

Without loss of generality, one can always subtract every point from its centroid to obtain
another  {\bf zero-mean} dataset $\tilde{\Omega}=\{x_1, x_2, \ldots, x_n\}\in\mathbb{R}^p$.
For such a zero-mean dataset, the cohesion measure of two points is simply the inner product between the two points, i.e.,
\begin{equation}
\gamma(x, y) = x^T y.
\end{equation}
Let
$$Q=\begin{bmatrix}
	x_1 & x_2 & \dots & x_n
\end{bmatrix}.$$
Then, the $n \times n$ semi-cohesion measure matrix
\begin{equation}
\Gamma_{n\times n}=
\begin{bmatrix}
    x_1^Tx_1 & x_1^Tx_2 & \dots & x_1^Tx_n \\
    x_2^Tx_2 & x_2^Tx_2 & \dots & x_2^Tx_n \\
    \vdots & \vdots & \ddots & \vdots \\
    x_n^Tx_1 & x_n^Tx_2 & \dots & x_n^Tx_n
\end{bmatrix}
=
\begin{bmatrix}
	x_1^T \\
	x_2^T \\
	\vdots \\
	x_n^T
\end{bmatrix}
\begin{bmatrix}
	x_1 & x_2 & \dots & x_n
\end{bmatrix}
= Q^TQ.
\end{equation}
For all $z\in\mathbb{R}^p$,
\begin{equation}
z^T\Gamma z = z^TQ^TQz = (Qz)^T(Qz) \geq 0.
\end{equation}
Thus, the semi-cohesion matrix obtained from using the {\em squared} Euclidean distance is a positive semi-definite matrix.
As such, there are $n$ nonnegative eigenvalues.
Let $\lambda_1\geq\lambda_2\geq\ldots\geq\lambda_n\geq0$ be the $n$ nonnegative eigenvalues of $\Gamma$, and $v_i$ be the corresponding (normalized) eigenvector for $\lambda_i$ with
 $\|v_i\|=1$. Thus, for $i=1,2, \ldots, n$,
 $$\Gamma v_i=\lambda_i v_i.$$
Moreover, these $n$ eigenvectors are orthogonal, i.e.,
 $v_i^T\cdot v_j=0$, for all  $i\neq j$.
Now let $V$ be the $n \times n$ matrix with $i^{th}$ column being the $i^{th}$ eigenvector, i.e.,
$$ V=\begin{bmatrix}
	v_1 & v_2 & \dots & v_n
\end{bmatrix}
$$
and $D$ be the $n \times n$ diagonal matrix with the $i^{th}$ diagonal element being the $i^{th}$ eigenvalue, i.e.,
$$D=\begin{bmatrix}
	\lambda_1 & \dots & \dots & 0 \\
	0 & \lambda_2 & \dots & 0\\
	\vdots & \vdots & \ddots & \vdots \\
	0 & 0 & \dots & \lambda_n
\end{bmatrix}.
$$
Thus,
\bear{pca2222}
&&\Gamma V\nonumber \\
&=&\Gamma
\begin{bmatrix}
	v_1 & v_2 & \dots & v_n
\end{bmatrix}\nonumber \\
&=&
\begin{bmatrix}
	\lambda_1 v_1 & \lambda_2 v_2 & \dots & \lambda_n v_n
\end{bmatrix}
\nonumber \\
 &=&
\begin{bmatrix}
	v_1 & v_2 & \dots & v_n
\end{bmatrix}
\begin{bmatrix}
	\lambda_1 & \dots & \dots & 0 \\
	0 & \lambda_2 & \dots & 0\\
	\vdots & \vdots & \ddots & \vdots \\
	0 & 0 & \dots & \lambda_n
\end{bmatrix}\nonumber \\
&=&VD.
\eear
As the matrix $V$ is an orthogonal matrix, its inverse $V^{-1}$ is simply its transpose matrix, i.e.,
$V^{-1}=V^T$. In view of \req{pca2222}, we then have
\beq{pca3333}
\Gamma = VDV^T = \sum\limits_{i=1}^n \lambda_i v_i^T v_i .
\eeq

Next, we assume that there is a low dimensionality space $\mathbb{R}^{p_0}$ and its semi-cohesion measure matrix can be denoted by
\begin{equation}
\tilde{\Gamma} = \sum\limits_{k=1}^{P_0}\lambda_k v_k v_k^T
\end{equation}
where
\begin{equation}
v_k =
\begin{bmatrix}
	v_{k, 1} \\
	v_{k, 2} \\
	\vdots \\
	v_{k, n}
\end{bmatrix}.
\end{equation}
Then,
\begin{equation}
\begin{aligned}
\tilde{\Gamma}_{ij} &= \sum\limits_{k=0}^{p_0}\lambda_k v_{k, i} v_{k, j} \\
&= \sum\limits_{k=0}^{p_0}(\sqrt{\lambda_k}v_{k, i})(\sqrt{\lambda_k}v_{k, j}) \\
&= (x'_i)^T(x'_j)
\end{aligned}
\end{equation}
and
\begin{equation}
x'_i =
\begin{bmatrix}
	\sqrt{\lambda_1}v_{1, i} \\
	\sqrt{\lambda_2}v_{2, i} \\
	\sqrt{\lambda_3}v_{3, i} \\
	\vdots \\
	\sqrt{\lambda_{p_0}}v_{p_0, i}
\end{bmatrix}.
\end{equation}
Thus, the eigendecomposition of the semi-cohesion matrix with the squared Euclidean distance is equivalent to using the principal component analysis (PCA) to map $x_i\in\mathbb{R}^{p}$ from a high-dimensional space  to $x'_i\in\mathbb{R}^{p_0}$ in a low-dimensional space.

\subsection{Using the softmax clustering algorithm with a hierarchical agglomerative  clustering}

There are two drawbacks of the softmax clustering algorithm: (i) the output of the algorithm may not be a cluster that satisfies the definition of a cluster in \rdef{community}, and (ii) the dataset may exist a hierarchical structure of clusters. To address these two drawbacks, one can use the same approach as described in the iterative Partitional-Hierarchical community Detection ($i$PHD) algorithm in  \cite{chang2016}. The key idea is to add a hierarchical agglomerative clustering after the softmax clustering algorithm. The hierarchical agglomerative  clustering algorithm then repeatedly  merges two positively correlated  clusters (produced by the softmax algorithm) into a new cluster until there is only one cluster left or every pair of two clusters are negatively correlated. The output clusters from such a hierarchical agglomerative  clustering can be shown to satisfy the definition of a cluster in \rdef{community}. Moreover, it also allows us to see the hierarchical structure of clusters. The detailed algorithm is shown in Algorithm \ref{alg:iphd}.

In the iterative partitional-hierarchical algorithm in Algorithm \ref{alg:iphd}, the modularity is non-decreasing when there is a change of the partition. Thus, the algorithm converges to a local optimum of the modularity in a finite number of steps. When the algorithm converges, every set returned by Algorithm \ref{alg:iphd} is indeed a cluster.

\begin{algorithm}[t]
\KwIn{A symmetric covariance matrix $\Gamma=(\gamma_{ij})$, the number clusters $K$, the inverse temperature $\theta>0$, and the annealing parameter $\epsilon\geq 0$.
 }
\KwOut{A partition ${\cal P}=\{S_k, k=1,2, \ldots, K_0\}$ with some $K_0 \le K$, and  an embedding of data points $\{z_i(k),i=1,2,\ldots,n,\;k=1,2,\ldots, K_0\}$.}

\noindent {\bf (0)} Initially, choose arbitrarily $K$ disjoint nonempty sets $S_1 , \ldots, S_K$ as a partition ${\cal P}$ of $\Omega$.

\noindent {\bf (1)} Run the softmax clustering algorithm in Algorithm \ref{alg:Softmax} with the initial partition ${\cal P}$. Let ${\cal P}^\prime$ be its output partition.

\noindent {\bf (2)} Run the hierarchical agglomerative  algorithm in \cite{chang2016} with the initial partition ${\cal P}^\prime$. Let ${\cal P}$ be its output partition.

\noindent {\bf (3)} Repeat from (1) until there is no further change of the partition.

\noindent {\bf (4)} Output the partition and the corresponding embedding  $\{z_i(k),i=1,2,\ldots,n,\;k=1,2,\ldots, K_0\}$ with
$$z_i(k)=\sum_{j \in S_k}\gamma_{j,i}.$$

\caption{The Iterative Partitional-Hierarchical ($i$PHD) Algorithm}
\label{alg:iphd}
\end{algorithm}

\begin{figure}[htbp]
	\centering
	\begin{tabular}{p{0.47\textwidth}p{0.47\textwidth}}
		\includegraphics[width=0.44\textwidth]{BlackGT.eps} &
		\includegraphics[width=0.44\textwidth]{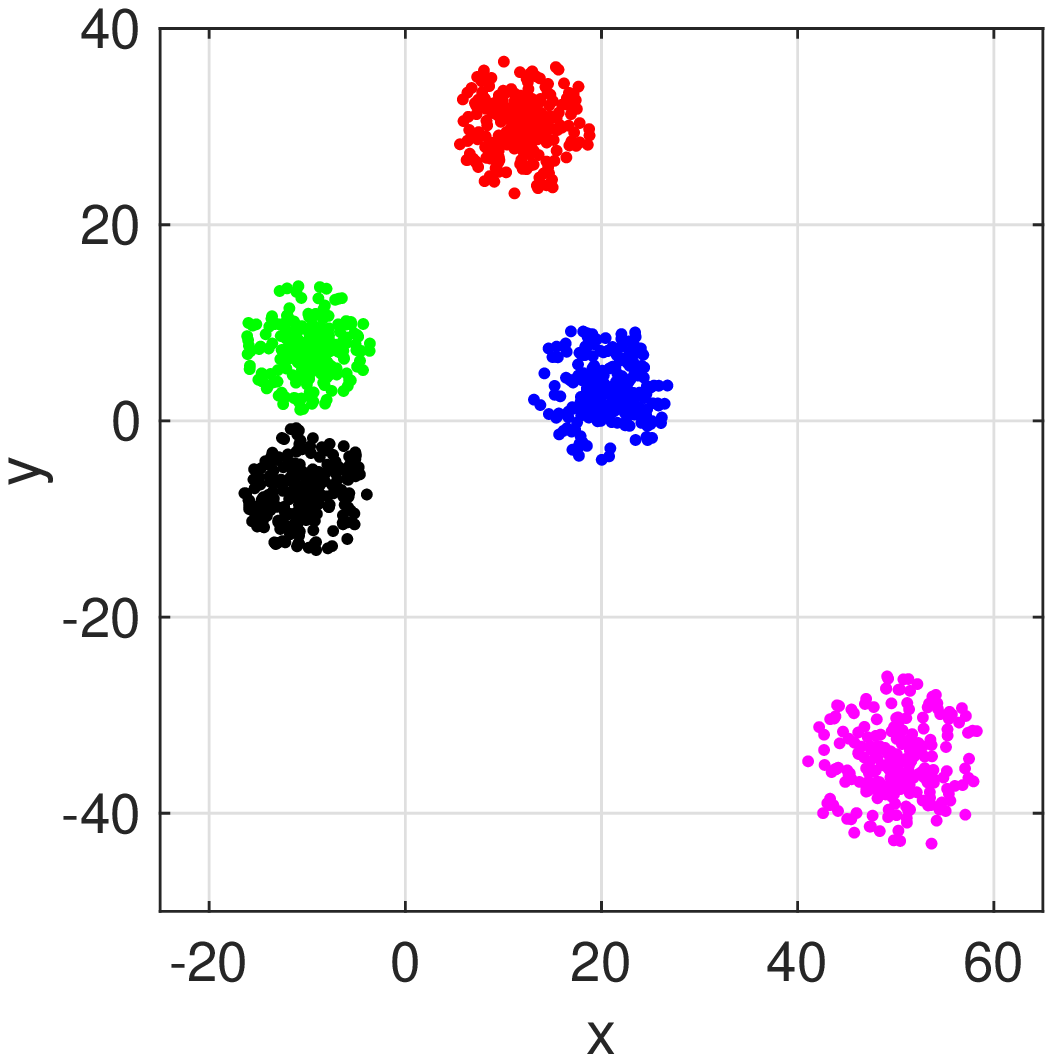} \\
		(a) The five-circle dataset & (b) The five clusters with $\lambda=-0.5$ \\
        \includegraphics[width=0.44\textwidth]{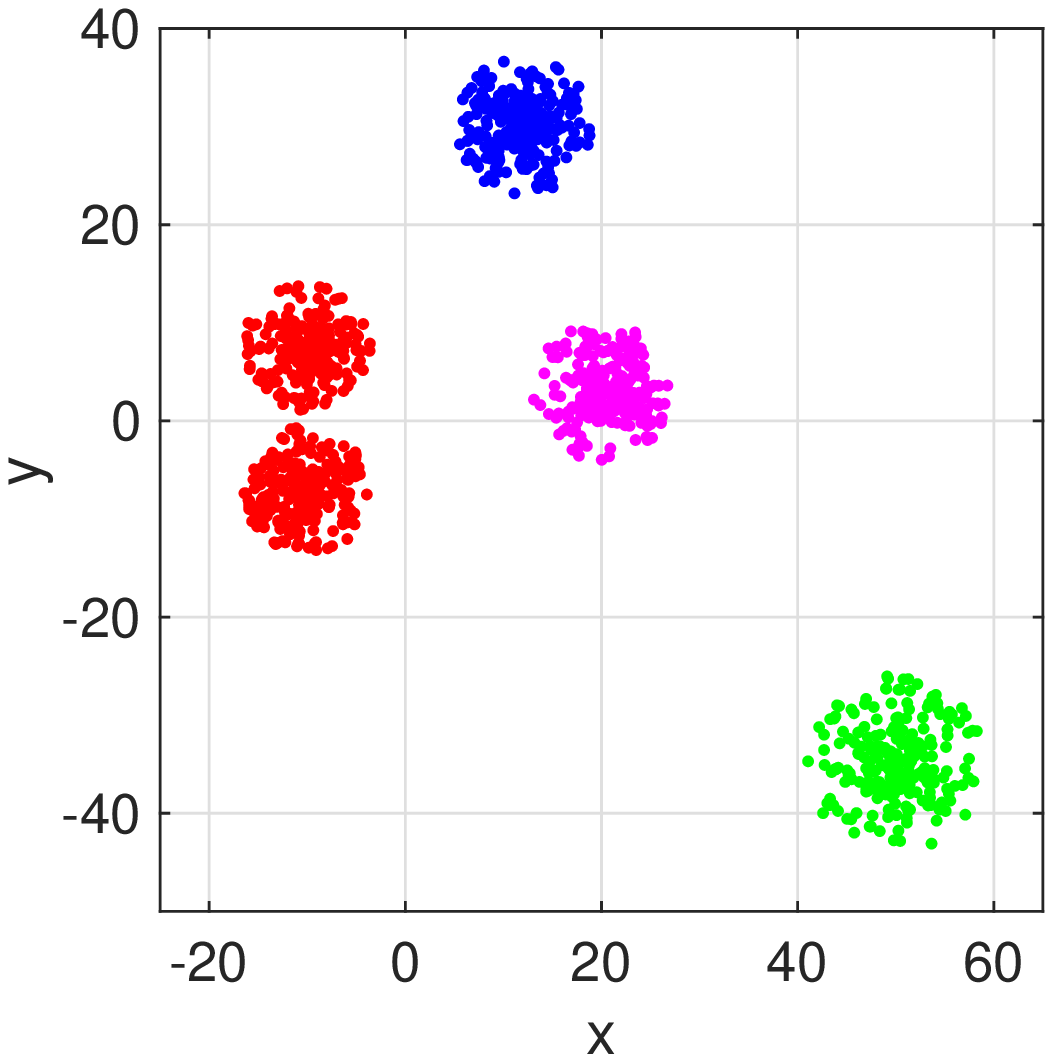} &
		\includegraphics[width=0.44\textwidth]{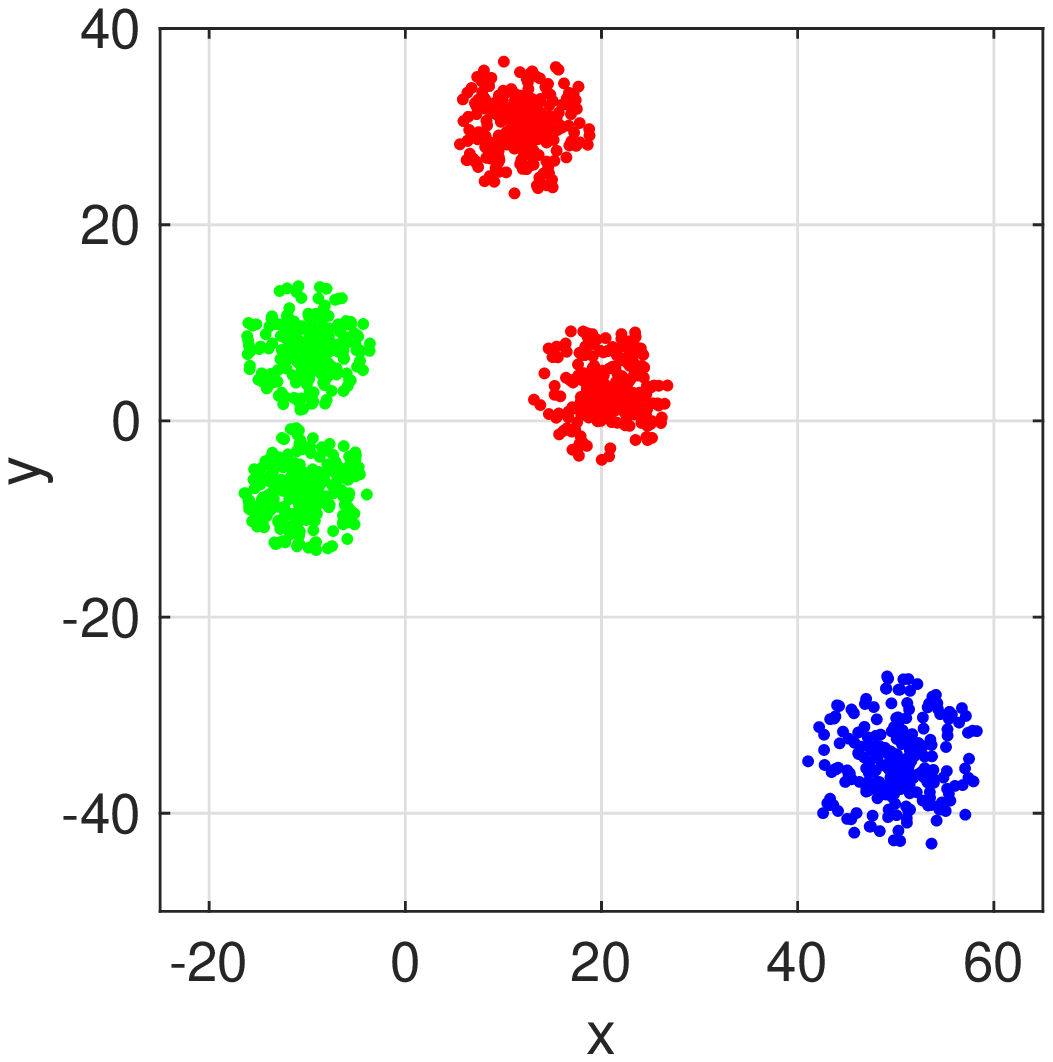} \\
		(c) The four clusters with $\lambda=-0.01$ & (d) The three clusters with $\lambda=-0.0001$ \\
	\end{tabular}
	\caption{An illustrating example for various resolutions of $\bar d$ (points in different clusters are marked with different colors).}
	\label{fig:fivecircleAll}
\end{figure}

As an illustrating example of the $i$PHD Algorithm, we consider the five-circle dataset in \rfig{fivecircle} (a). As shown in \rfig{fivecircle} (b), one can select a particular  $\lambda$'s so that the average distance is $\bar d$ under the sampling distribution $p_\lambda(\cdot,\cdot)$. In Table \ref{tab:lambdadbar}, we list the average distance $\bar d$ for various choices of $\lambda$. Now we feed the covariance matrix from the sampling distribution $p_\lambda(\cdot,\cdot)$ into the $i$PHD Algorithm. It is clear to see from \rfig{fivecircleAll} that various choices of $\lambda$ lead to various resolutions of the clustering algorithm. Specifically, for $\lambda=-0.5$ (and $\bar d=2.6055$), there are five clusters detected by the $i$PHD algorithm. Points in different clusters are marked with different colors. For  $\lambda=-0.01$ (and $\bar d=31.1958$), there are four clusters detected by the $i$PHD algorithm. Finally, for  $\lambda=-0.0001$ (and $\bar d=36.6545$), there are only three clusters detected by the $i$PHD algorithm.

\begin{table*}[ht]
\centering
\caption{The average distance $\bar d$ for various choices of $\lambda$.\label{tab:lambdadbar}}
\begin{tabular}{|c||c|c|c|c|c|}
\hline
$\lambda$ & -0.5 &	-0.01&	-0.0001 &	0	&1 \\ \hline
$\bar d$ & 2.6055	&31.1958	&36.6545	&36.7121	& 87.8800 \\ \hline
\end{tabular}
\end{table*}

\section{Performance comparisons}

\subsection{Choice of the objective function}

In this section, we compare the performance of clustering algorithms that use different objective functions. Our softmax clustering algorithm and its $i$PHD extension use the modularity $\sum_{k=1}^K \gsim(S_k,S_k)$ as its objective function. On the other hand, the  K-sets$^+$ algorithm in \cite{ksetsplus} uses the {\em normalized} modularity $\sum_{k=1}^K \frac{1}{|S_k|}\gsim(S_k,S_k)$ as its objective function.

Here we briefly review the K-sets$^+$ algorithm. In \cite{ksets}, a clustering algorithm, called the K-sets algorithm, was proposed for clustering  data points in metric spaces. The K-sets algorithm is conceptually simple, and it relies on a new distance measure, called the {\em triangular distance ($\Delta$-distance)} in \cite{ksets}. The K-sets algorithm, started from a random partition of $K$ sets, repeatedly assigns every data point to the closest set in terms of the $\Delta$-distance until there is no further change of the partition. The K-sets algorithm was extended to the K-sets$^+$ algorithm in \cite{ksetsplus} for clustering data points in semi-metric spaces. The key problem for such an extension is that the {\em triangular distance ($\Delta$-distance)} might not be nonnegative in a semi-metric space. For this, the K-sets$^+$ algorithm uses the adjusted  $\Delta$-distance defined below.

 \bdefin{ADeltaD}{\bf (Adjusted  $\Delta$-distance)} For a semi-cohesion measure $\gsim(\cdot,\cdot)$ on a set of data points $\Omega=\{x_1, x_2, \ldots, x_n\}$,
the {\em adjusted  $\Delta$-distance}   from a point $x$ to a set $S$, denoted by $\Deltaw(x, S)$, is defined as follows:
\begin{equation}
\Deltaw(x,S)=\left \{
\begin{array}{ll}
\frac{|S|}{|S|+1} \Delta(x,S), & \mbox{if}\;x \not \in S, \\
\frac{|S|}{|S|-1} \Delta(x,S), & \mbox{if}\;x \in S\;\mbox{and}\; |S|>1,\\
-\infty,& \mbox{if}\; x \in S\;\mbox{and}\;|S|=1.
                \end{array} \right. ,
\end{equation}
where
\begin{equation}
\Delta(x, S)=\gsim(x,x)-\frac{2}{|S|} \gsim(x,S)+ \frac{1}{|S|^2}\gsim(S,S),
\end{equation}
\edefin


Instead of using the $\Delta$-distance for the assignment of a data point in the K-sets algorithm, the K-sets$^+$ algorithm uses the adjusted $\Delta$-distance for the assignment of data points. It was shown in \cite{ksetsplus} that the K-sets$^+$ algorithm can also be used for clustering data points with a symmetric similarity measure (that measures how similar two data points are) and it converges monotonically to a local optimum of the optimization problem for the objective function $\sum_{k=1}^K \frac{1}{|S_k|}\gsim(S_k,S_k)$ within a finite number of iterations. The detailed algorithm is shown in Algorithm \ref{alg:ksetsplus}.



\begin{algorithm}[t]
\KwIn{A data set $\Omega=\{x_1, x_2, \ldots, x_n\}$, a {\em symmetric} matrix $\Gamma=(\gsim(\cdot, \cdot))$ and the number of sets $K$.
}
\KwOut{A partition of sets $\{S_1, S_2, \ldots, S_K\}$.
}

\noindent {\bf (0)} Initially, choose arbitrarily $K$ disjoint nonempty sets $S_1 , \ldots, S_K$ as a partition of $\Omega$.

\noindent {\bf (1)} \For{$i=1, 2, \ldots, n$}{

\noindent
Compute the adjusted $\Delta$-distance $\Deltaw(x_i, S_k)$ for each set $S_k$.
Find the set to which the point $x_i$ is closest in terms of the adjusted $\Delta$-distance. Assign that point $x_i$ to that set.}

\noindent {\bf (2)} Repeat from (1) until there is no further change.
\caption{The \ksetsplus Algorithm \cite{ksetsplus}}
\label{alg:ksetsplus}
\end{algorithm}

To study the effect of the choice of the objective function, we compare the $i$PHD algorithm with the K-sets$+$ algorithm. Recall that the $i$PHD algorithm uses the modularity $\sum_{k=1}^K \gsim(S_k,S_k)$ as its objective function and the K-sets$^+$ algorithm  uses the {\em normalized} modularity $\sum_{k=1}^K \frac{1}{|S_k|}\gsim(S_k,S_k)$ as its objective function. For our experiments, we follow the procedure in \cite{CD_ECC} to generate signed networks from the stochastic block model with two communities.  Each sign network consists of $n$ nodes and two ground-truth blocks. However, the sizes of these two blocks are chosen differently. There are three key parameters $p_{in}$, $p_{out}$, and $p$ for generating a test network. The parameter $p_{in}$ is the probability that there is a \textit{positive} edge between two nodes within the same block and $p_{out}$ is the probability that there is a \textit{negative} edge between two nodes in two different blocks. All edges are generated independently according to $p_{in}$ and $p_{out}$. After all the signed edges are generated, we then flip the sign of an edge independently with the crossover probability $p$.

In our experiments, the total number of nodes is $n=$ 2000 with four choices of $n_1=\{$1000, 1300, 1600, 1900$\}$  in the first block (and thus $n_2=\{$1000, 700, 400, 100$\}$ in the second block). Let $c=(n/2-1)p_{in}+np_{out}/2$ be the average degree of a node, and it is set to be 6, 8, and 10, respectively. Also, let $c_{in}=np_{in}$ and $c_{out}=np_{out}$. The value of $c_{in}-c_{out}$ is set to be 5 and that is used with the average degree $c$ to uniquely determine $p_{in}$ and $p_{out}$. The crossover probability $p$ is in the range from 0.01 to 0.5 with a common step of 0.01. We generate 20 graphs for each $p$ and $c$. We remove isolated nodes, and thus the exact numbers of nodes in the experiments might be less than 2000. We show the experimental results with each point averaged over 20 random graphs. The error bars represent the 95\% confident intervals.

To test these two algorithms, we use the similarity matrix $\Gamma$ with
\begin{equation}
\label{eq:AA2}
\Gamma = A+0.5A^2,
\end{equation}
where $A$ is the adjacency matrix of the signed network after randomly flipping the sign of an edge. Such a similarity matrix was suggested in \cite{CD_ECC} for community detection in signed networks as it allows us to ``see'' more than one step relationship between two nodes.

\begin{figure}[htbp]
	\includegraphics[width=1\textwidth]{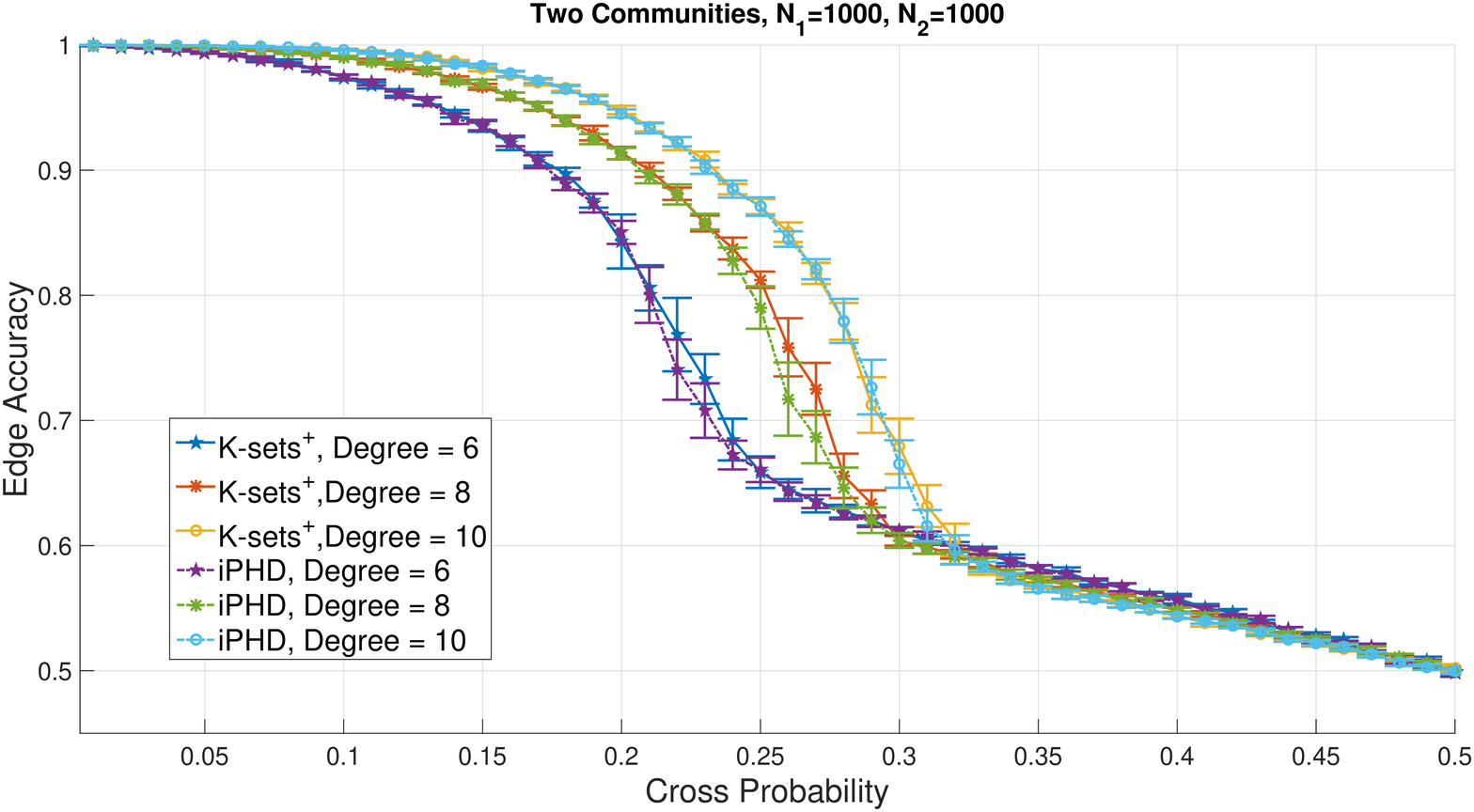}
	\caption{The stochastic block model with 1000 and 1000 nodes in each community.}
	\label{fig:1000_1000}
\end{figure}

\begin{figure}[htbp]
	\includegraphics[width=1\textwidth]{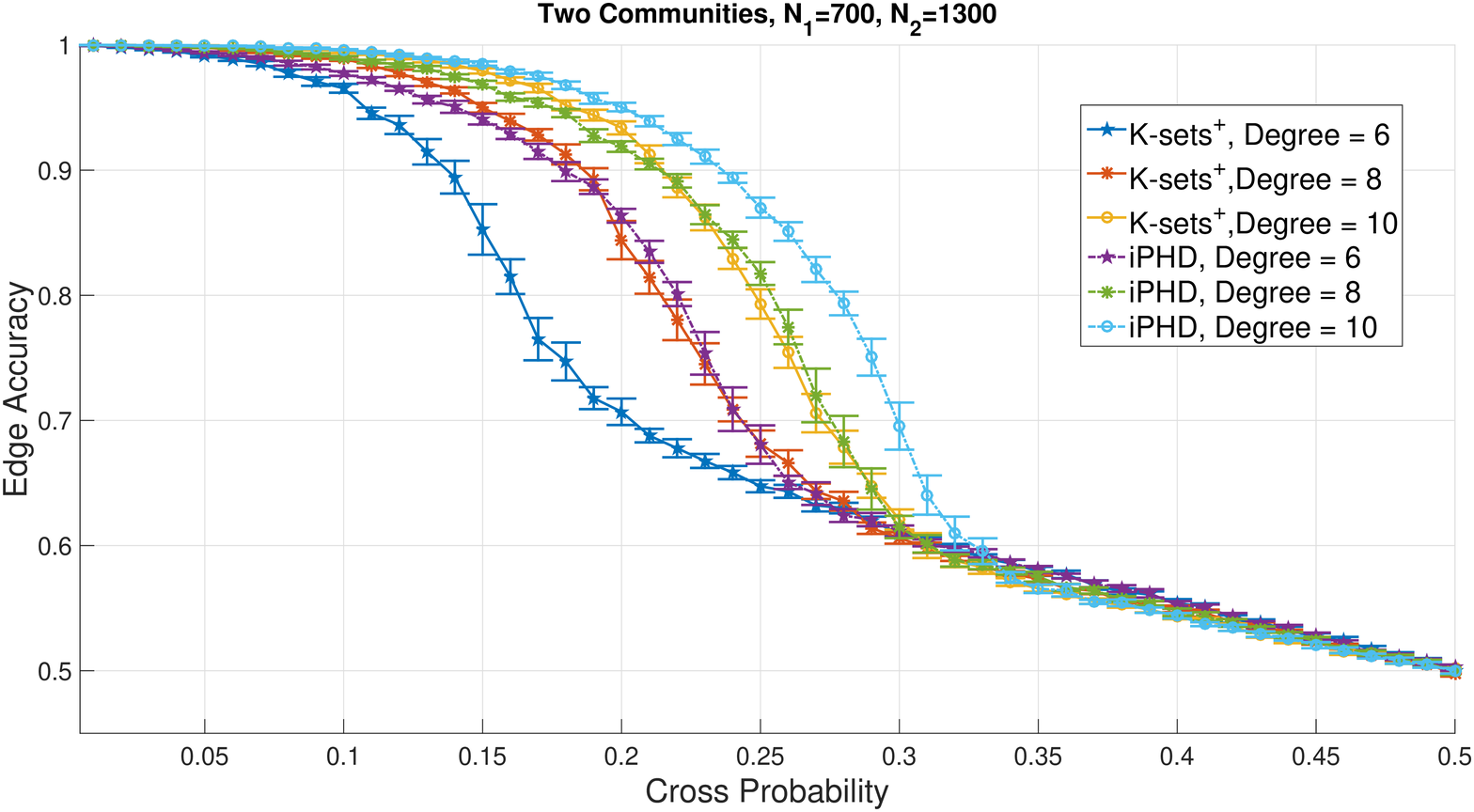}
	\caption{The stochastic block model with 700 and 1300 nodes in each community.}
	\label{fig:700_1300}
\end{figure}

\begin{figure}[htbp]
	\includegraphics[width=1\textwidth]{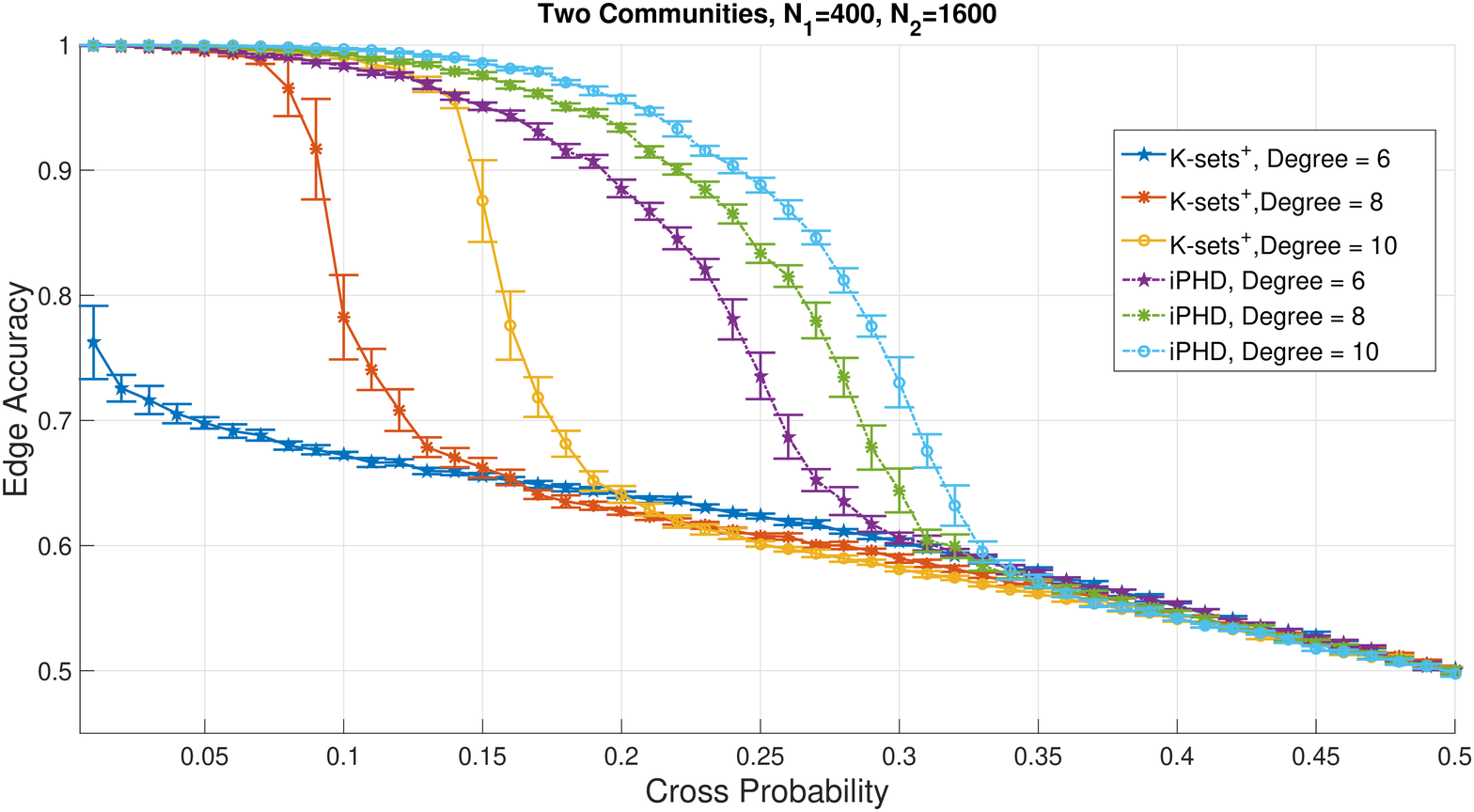}
	\caption{The stochastic block model with 400 and 1600 nodes in each community.}
	\label{fig:400_1600}
\end{figure}

\begin{figure}[htbp]
	\includegraphics[width=1\textwidth]{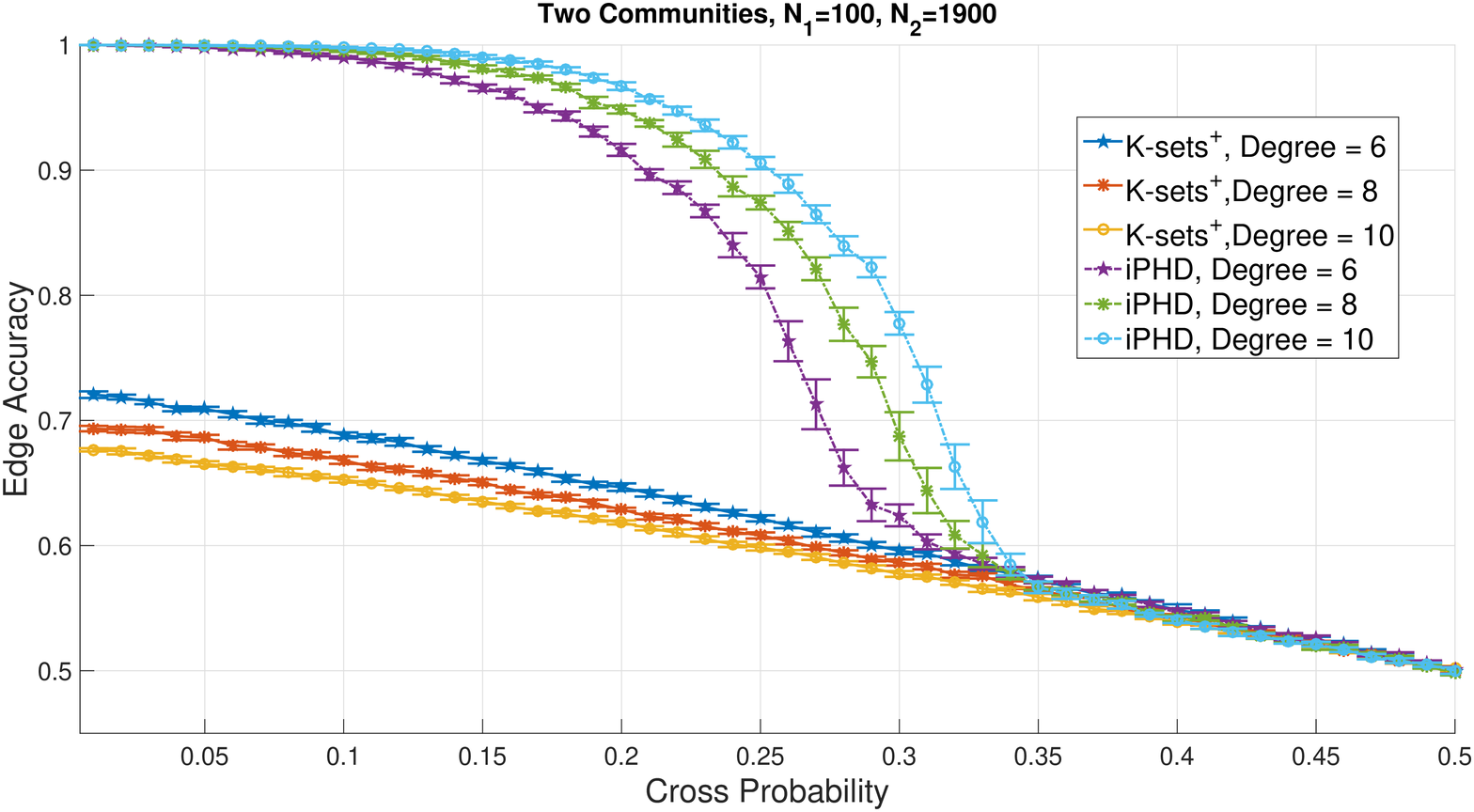}
	\caption{The stochastic block model with 100 and 1900 nodes in each community.}
	\label{fig:100_1900}
\end{figure}

In \rfig{1000_1000}, \rfig{700_1300}, \rfig{400_1600}, and \rfig{100_1900}, we show the experimental results for edge accuracy (the percentage of edges that are correctly detected) as a function of the crossover probability $p$.

From our experimental results, we can observe that the K-sets$^+$ algorithm does not perform well when the sizes of communities are  considerably different. The reason is that the normalized modularity tends to balance the sizes of the two detected communities. Also, increasing the average degree $c$ in the stochastic block model also increases the edge accuracy for both algorithms. This might be due to the fact that the tested signed networks with a larger average degree are more dense. Thus, if we would like to cluster data with communities of the same size, we should use the K-sets$^+$ (that maximizes the normalized modularity) to obtain more precise results. On the contrary, if the ground-truth communities are not of the same size, we should use the $i$PHD algorithm (that maximizes the modularity).

To further verify the above insight, we test the $i$PHD algorithm and the K-sets$^{+}$ algorithm on the real-world dataset from the LiveJournal \cite{LiveJournal1,LiveJournal2}. LiveJournal is a free on-line community with almost 10 million members in which  a significant fraction of members are highly active. In our experiment, we extract the largest two non-overlapping communities. The total number of nodes  $n$ is  2583 with 1243 and 1340 nodes in each community. The crossover probability $p$ is in the range from 0.01 to 0.25 with a common step of 0.01 and we use the same similarity matrix $\gsim$ in \req{AA2}.

\begin{figure}[htbp]
	\includegraphics[width=1\textwidth]{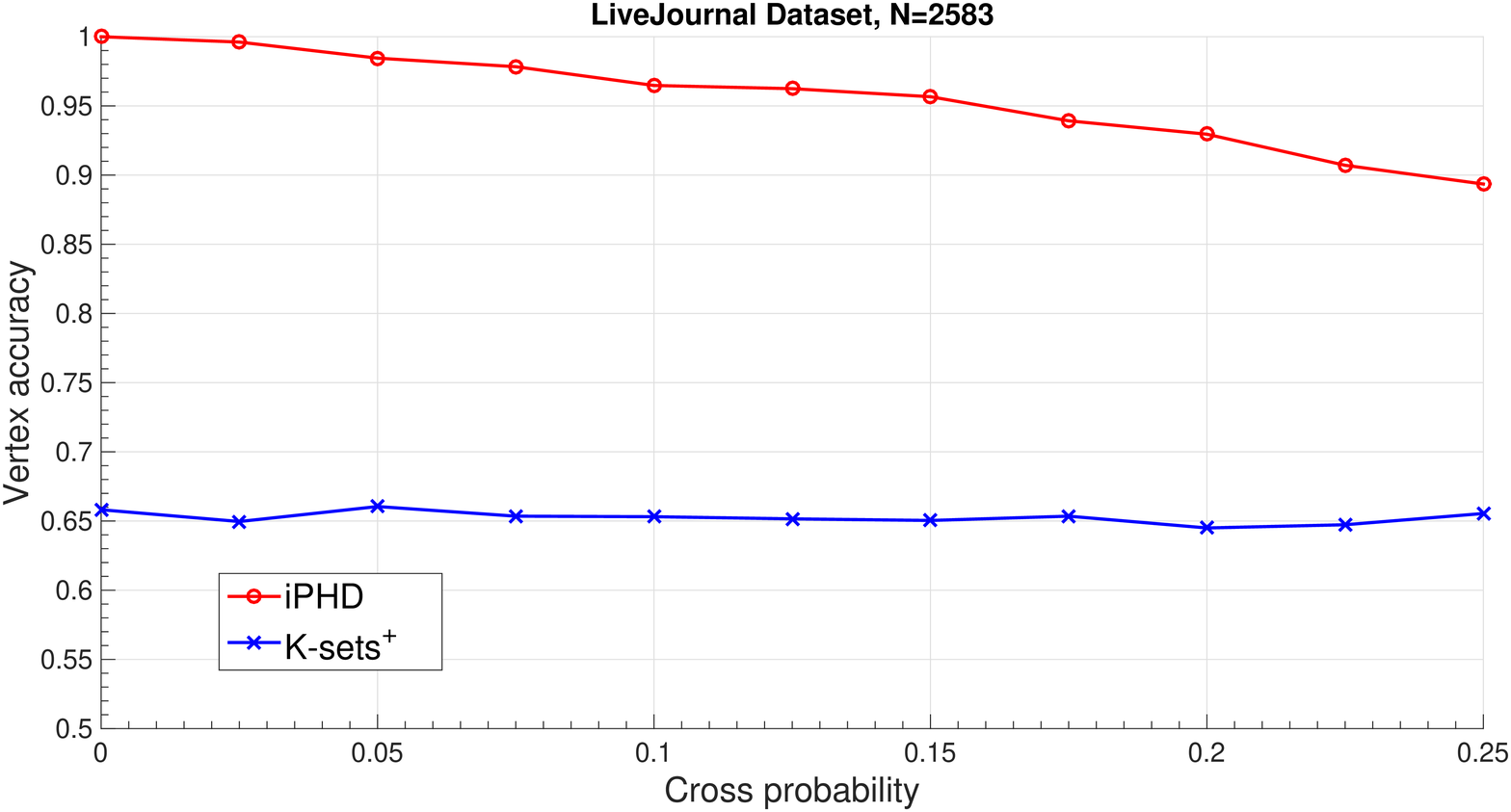}
	\caption{The comparison results for the K-set$+$ algorithm and the $i$PHD algorithm by using the LiveJournal dataset.}
	\label{fig:LiveJournal}
\end{figure}

In \rfig{LiveJournal}, we show our experimental results for vertex accuracy (the percentage of vertices that are correctly clustered) as a function of the crossover probability $p$. As shown in \rfig{LiveJournal}, the K-sets$^+$ do not perform very well. However, it achieves a higher objective value (in terms of the normalized modularity) than that of the ground-truth communities. This is very interesting as the K-sets$^+$ algorithm does its job to produce communities with high normalized modularity values. But this does not imply that these communities with high normalized modularity values are close the ground-truth communities. We will further discuss this in the next section.

\subsection{Choice of the similarity/dissimilarity measure}

In the previous section, we show that it is possible for a clustering algorithm to produce communities/clusters with higher objective values than that of the ground-truth communities and they are not even close to the ground-truth communities. Such an observation makes us wonder whether one can fix this problem by using a ``right'' similarity measure. It is known in \cite{ksetsplus} that one can convert a similarity measure $\gsim(x, y)$ into a semi-cohesion measure $\tilde{\gsim}(x, y)$ as follows:
\beq{dissim0000}
\tilde{\gsim}(x, y)=\gsim(x, y)-\frac{1}{n}\gsim(x, \Omega)-\frac{1}{n}\gsim(y, \Omega)+\frac{1}{n^2}\gsim(\Omega, \Omega)+\sigma\delta(x, y)-\frac{\sigma}{n},
\eeq
where $\delta(x, y)$ is the usual $\delta$ function (that has value 1 if $x=y$ and 0 otherwise), and $\sigma$ is a constant that satisfies
\begin{equation}
\sigma\geq \max\limits_{x\neq y}[\gsim(x, y)-(\gsim(x, x)+\gsim(y, y))/2].
\end{equation}
Once we have the semi-cohesion measure, we can use the duality result in \req{cind1111} to construct a semi-metric
\beq{dissim1111}
\tilde d(x,y)=(\tilde \gsim(x,x)+\tilde \gsim(y,y))/2 -\tilde \gsim(x,y).
\eeq
To convert such a semi-metric into a metric $\tilde d^*(x,y)$, we can use  the shortest-path algorithm, e.g., the Dijkstra algorithm \cite{Dijkstra}, to compute the minimum distance between $x$ and $y$. Clearly, the minimum distance satisfies the triangular inequality and thus one can use the K-sets algorithm in \cite{ksets} with the distance metric $\tilde d^*(x,y)$.

\begin{figure}[htbp]
	\centering
	\includegraphics[width=1\textwidth]{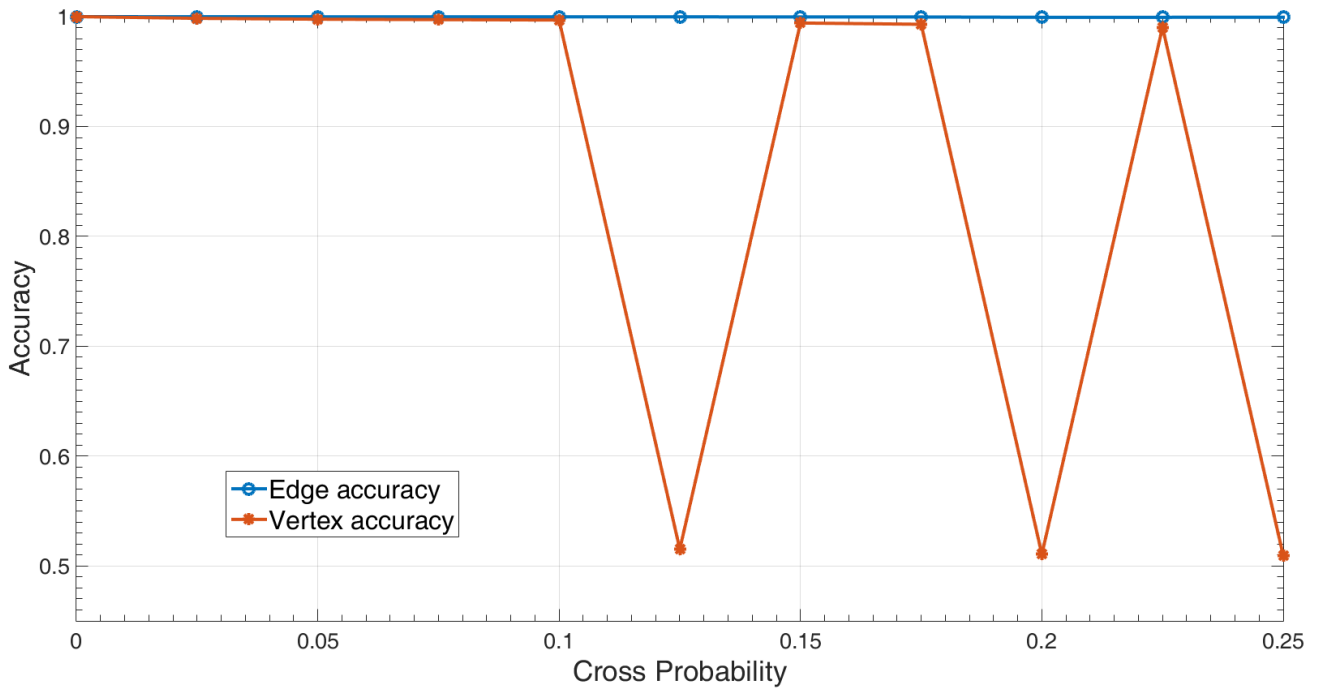}
	\caption{The result for  the K-sets algorithm  by using the LiveJournal dataset.}
	\label{fig:KsetsLiveJournal}
\end{figure}

In \rfig{KsetsLiveJournal}, we show the  experimental result for vertex and edge accuracy for the LiveJournal dataset \cite{LiveJournal1,LiveJournal2} by using the K-sets algorithm in \cite{ksets} with the distance metric $\tilde d^*(x,y)$. Each point in \rfig{KsetsLiveJournal} is the best result in 500 tries of the K-sets algorithm with random initialization. As shown in \rfig{KsetsLiveJournal}, both the performance for edge accuracy and that of vertex accuracy are good except for the three abnormal results at $p=0.125$, $p=0.20$ and $p=0.25$ as there are only two negative edges between the two communities in this dataset. If the edge signs of these two edges are flipped, then we might cluster most of the nodes into a wrong cluster. Thus, even though the total number of  edge errors is still two, the number of vertex errors is tremendous. In comparison with the results from the K-sets$+$ algorithm in \rfig{LiveJournal}, there is a significant performance improvement by using the K-sets algorithm that uses the distance metric $\tilde d^*(x,y)$. This shows that the choice of the distance measure has a great impact on the performance of the clustering algorithm.

\begin{figure}[htbp]
	\centering
	\includegraphics[width=0.5\textwidth]{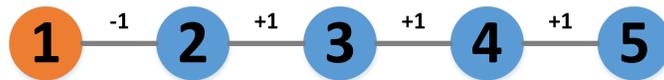}
	\caption{A line graph with 5 nodes}
	\label{fig:cex1}
\end{figure}

To gain the intuition why using a metric is better than using a semi-metric, let us consider a simple line graph with 5 nodes as shown in \rfig{cex1}. There is only one negative edge, i.e., the edge between node 1 and 2, and the rest three edges are positive edges. Intuitively, such a signed network should be clustered into two communities $\{1\}$ and $\{2,3,4,5\}$.
Note that the adjacency matrix $A$ for the signed network is
$$A=\begin{bmatrix}
	0 & -1 & 0 & 0  &0 \\
	-1 & 0 & 1 & 0  &0 \\
    0 &  1 & 0 & 1  &0 \\
	0 & 0 &  1 & 0 & 1  \\
	0 &  0 & 0 & 1  &0 \\
\end{bmatrix}.
$$
Now treat the adjacency matrix $A$ as the similarity measure $\gsim(\cdot,\cdot)$ and use \req{dissim0000} and \req{dissim1111} to convert such a similarity measure into a semi-metric. This leads to
\beq{dissim3333}
\tilde d=\begin{bmatrix}
	0 & 2 & 1 & 1  &1 \\
	2& 0 & 0 & 1  &1 \\
    1 &  0 & 0 & 0  &1 \\
	1 & 1 &  0 & 0 & 0  \\
	1 &  1 & 1 & 0  &0 \\
\end{bmatrix}.
\eeq
It is known in \cite{ksets} that the normalized modularity satisfies the following identity:
\beq{equiv1111}
\sum_{k=1}^K \frac{1}{|S_k|}\gsim(S_k,S_k)
=\sum_{i=1}^n \gsim(i,i)-\sum_{k=1}^K \frac{1}{|S_k|}{\tilde d}(S_k,S_k).
\eeq
Thus, maximizing the normalized modularity  is equivalent to minimizing $\sum_{k=1}^K \frac{1}{|S_k|}{\tilde d}(S_k,S_k)$.
For the ground-truth communities $\{1\}$ and $\{2,3,4,5\}$, we have
$$\sum_{k=1}^K \frac{1}{|S_k|}{\tilde d}(S_k,S_k)=0+\frac{6}{4}=\frac{3}{2}.$$
However, for the partition of the two  communities $\{1,4,5\}$ and $\{2,3\}$, we have
$$\sum_{k=1}^K \frac{1}{|S_k|}{\tilde d}(S_k,S_k)=\frac{4}{3}+0=\frac{4}{3},$$
which is smaller than $3/2$ for the ground-truth communities. This example shows that maximizing the normalized modularity with a semi-metric may lead to a partition that has a higher objective value than that of the ground-truth communities.

Now let us convert the semi-metric in \req{dissim3333} into a metric by computing the minimum distance between any two points and this leads to
\beq{dissim4444}
\tilde d^*=\begin{bmatrix}
	0 & 1 & 1 & 1  &1 \\
	1& 0 & 0 & 0  &0 \\
    1 &  0 & 0 & 0  &0 \\
	1 & 0 &  0 & 0 & 0  \\
	1 &  0 & 0 & 0  &0 \\
\end{bmatrix}.
\eeq
For the ground-truth communities $\{1\}$ and $\{2,3,4,5\}$,
$$\sum_{k=1}^K \frac{1}{|S_k|}{\tilde d}(S_k,S_k)=0,$$
and this is the best objective value.
Note that for the partition of the two  communities $\{1,4,5\}$ and $\{2,3\}$,
$$\sum_{k=1}^K \frac{1}{|S_k|}{\tilde d}(S_k,S_k)=\frac{4}{3}+0=\frac{4}{3}.$$

\begin{figure}[htbp]
	\centering
	\includegraphics[width=0.3\textwidth]{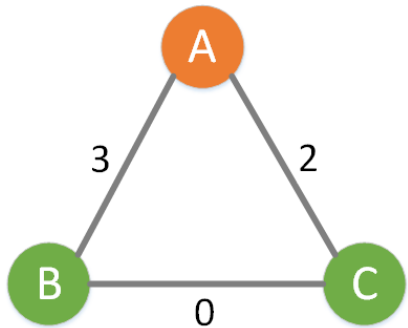}
	\caption{Three nodes with a semi-metric}
	\label{fig:ThreeNodes}
\end{figure}

The advantage of converting a semi-metric into a metric can be further explained from the three-node graph in \rfig{ThreeNodes}. There the triangular inequality is not satisfied. Since the distance between node B and node C is 0, it is intuitive to treat $B$ and $C$ as the same node. However, as the triangular inequality is not satisfied, node A sees them differently and that might cause misclustering of node B and node C.

\begin{figure}[htbp]
	\centering
	\begin{tabular}{p{0.48\textwidth}p{0.45\textwidth}}
		\includegraphics[width=0.40\textwidth]{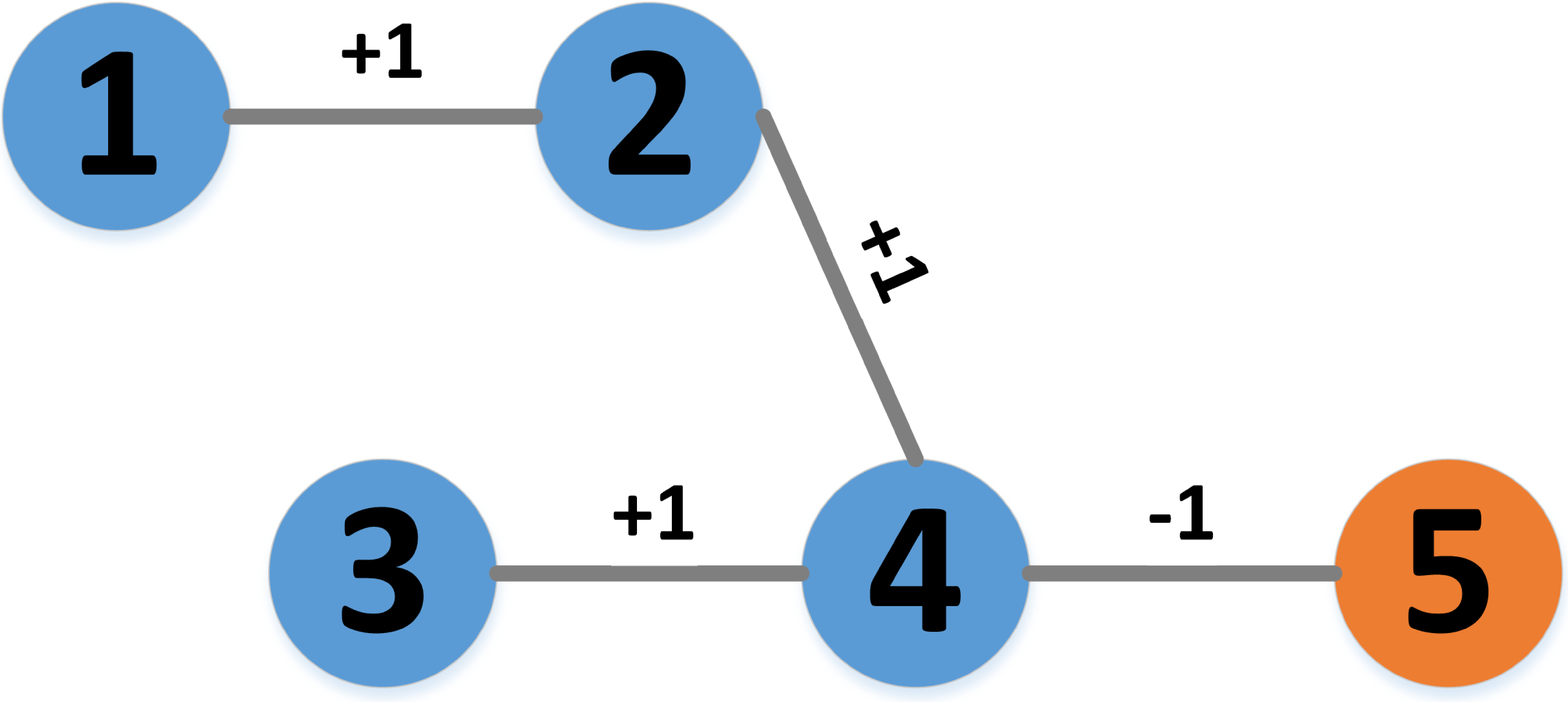} &
		\includegraphics[width=0.30\textwidth]{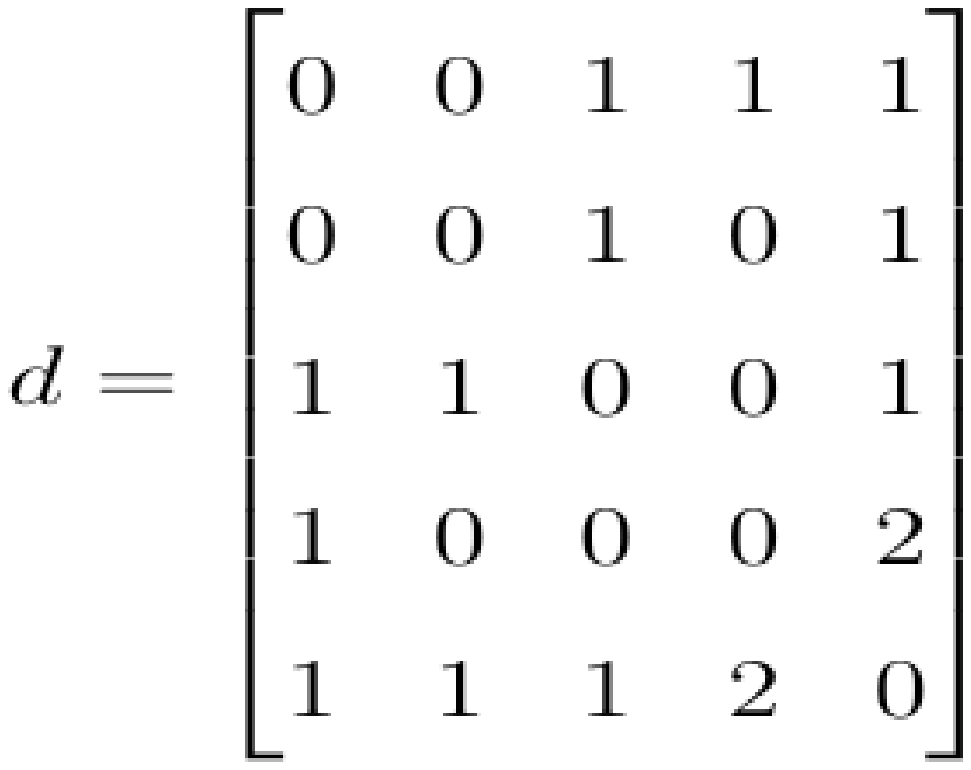} \\
		(a) a tree with 5 nodes & (b) semi-metric matrix\\
	\end{tabular}
	\caption{Another illustrating example for a tree with 5 nodes.}
	\label{fig:cex2}
\end{figure}

The line graph in \rfig{cex1} shows that it is possible for a clustering algorithm to produce communities/clusters with higher objective values than that of the ground-truth communities and they are not even close to the ground-truth communities. In \rfig{cex2}, we provide another example for a tree with 5 nodes. The only negative edge is the edge between node 4 and 5. Clearly, the ground-truth communities are $\{5\}$ and $\{1,2,3,4\}$. But for the partition with the two communities $\{1,2,5\}$ and $\{3,4\}$ has a larger normalized modularity than that of the  ground-truth communities.

\section{Conclusion}

In this paper, we proposed a unified framework for sampling, clustering and embedding data points in semi-metric spaces. We introduced the whole concept of clustering in a sampled graph with the exponentially twisted sampling. Then, we proposed a probabilistic clustering algorithm, called the softmax clustering algorithm based on the softmax function and the covariance for not only clustering but also embedding data points in a semi-metric space to a low dimensional Euclidean space. We showed that the softmax clustering algorithm converges to a local optimum when the inverse temperature $\theta$ is increased to infinity. To show the effect of the softmax clustering algorithm, we also conducted an illustrating experiment by using an artificial dataset with three non-overlapping rings. Furthermore, we provided supporting evidence by using the eigendecomposition of the semi-cohesion measure from artificial datasets and showed that the eigendecomposition of the semi-cohesion matrix with the squared Euclidean distance is equivalent to using the principal component analysis (PCA) to map from a high-dimensional space to in a low-dimensional space. Also, to address drawbacks of the softmax clustering algorithm, we extended the $i$PHD algorithm and showed that experimental results with various choices of $\lambda$ lead to various resolutions of the clustering algorithm.

Besides, we focused on (i) how the choice of the objective function and (ii) the choice of the similarity/dissimilarity measure affect the performance of the clustering results. In the first part, we followed the procedure in \cite{CD_ECC}, and experimental results showed that those algorithms based on the maximization of normalized modularity tend to balance the size of detected clusters. In the second part, we showed that using a metric is better than using a semi-metric as the triangular inequality is not satisfied for a semi-metric and that is more prone to clustering errors.

\bibliographystyle{IEEEtran}
\bibliography{bibliographyChiaTai}

\end{document}